\documentclass{aa}  
\usepackage{txfonts}
\usepackage{xspace}
\usepackage{graphicx}
\usepackage{natbib}
\usepackage{longtable}
\usepackage{adjustbox}
\usepackage{xcolor}
\usepackage[colorlinks = true,
            linkcolor = blue,
            urlcolor  = blue,
            citecolor = blue]{hyperref}

\def\alphaco{$\alpha_{\rm CO}$}
\def\arcsec{\hbox{$^{\prime\prime}$}}
\def\beq{\begin{equation}}

\def\distSFMS{$\Delta${\rm SFMS}}
\def\eeq{\end{equation}}
\def\faper{$f_{\rm ap}$}
\def\icoone{$I_{\rm CO(1-0)}$}
\def\icotwo{$I_{\rm CO(2-1)}$}
\def\kms{$\mathrm{km \, s^{-1}}$}
\def\lco{$L^\prime_{\rm CO}$}

\def\mhtwo{$M_{\rm H_2}$}
\def\Mstar{$M_{\star}$}
\def\Msun{$M_{\odot}$}
\def\riso{$r_{\rm 25}$}
\def\rtwoone{$R_{\rm 21}$}

\begin{document} 

   \title{CO-CAVITY project: Molecular gas and star formation in void galaxies}

   \author{
          M. I. Rodríguez
          \inst{1}
          \and
          U. Lisenfeld \inst{1, 2}
          \and
          S. Duarte Puertas \inst{1, 2, 3}
          \and
          D. Espada \inst{1,2}
          \and
          J. Domínguez-Gómez \inst{1}
          \and
          M. Sánchez-Portal \inst{4}
          \and
          A. Bongiovanni \inst{4}
          \and
          M. Alcázar-Laynez \inst{1}
          \and
          M. Argudo-Fernández \inst{1, 2}
          \and
          B. Bidaran \inst{1}
          \and
          S. B. De Daniloff \inst{1, 4}
          \and
          J. Falcón-Barroso \inst{6,7}
          \and
          E. Florido \inst{1,2}
          \and
          R. García-Benito \inst{8}
          \and
          A. Jimenez \inst{1}
          \and
          K. Kreckel \inst{9}
          \and
          R. F. Peletier \inst{11}
          \and
          I. Pérez \inst{1,2}
          \and
          T. Ruiz-Lara \inst{1,2}
          \and
          L. Sánchez-Menguiano \inst{1,2}
          \and
          G. Torres-Ríos\inst{1}
          \and
          P. Villalba-González \inst{13}
          \and
          S. Verley \inst{1,2}
          \and
          A. Zurita \inst{1,2}
   }

   \institute{ 
              Departamento de F\'{\i}sica T\'eorica y del Cosmos, Universidad de Granada, 18071 Granada, Spain 
              \email{monica-rodriguez@ugr.es}
         \and
             Instituto Carlos I de F\'{\i}sica T\'eorica y Computacional, Facultad de Ciencias, 18071 Granada, Spain
         \and D\'epartement de Physique, de G\'enie Physique et d'Optique, Universit\'e Laval, and Centre de Recherche en Astrophysique du Qu\'ebec (CRAQ), Québec, QC, G1V 0A6, Canada
        \and
             Institut de Radioastronomie Millimétrique (IRAM), Av. Divina Pastora 7, Local 20, 18012 Granada, Spain
        \and
             Department of Physics and Astronomy, University of British Columbia, Vancouver, BC V6T 1Z1, Canada
        \and 
             Instituto de Astrof\'{\i}sica de Canarias, V\'{\i}a L\'actea s/n, 38205 La Laguna, Tenerife, Spain
        \and 
             Departamento de Astrof\'{\i}sica, Universidad de La Laguna, 38200 La Laguna, Tenerife, Spain
        \and 
             Instituto de Astrof\'{\i}sica de Andaluc\'{\i}a - CSIC, Glorieta de la Astronom\'{\i}a s.n., 18008 Granada, Spain
        \and 
             Astronomisches Rechen-Institut, Zentrum f\"ur Astronomie der Universit\"at Heidelberg, 69120 Heidelberg, Germany
        \and 
             Instituto de Astrof\'{\i}sica, Facultad de F\'{\i}sica, Pontificia Universidad Cat\'olica de Chile, Av. Vicu\~na Mackenna 4860, Santiago, Chile
        \and 
             Kapteyn Astronomical Institute, University of Groningen, Landleven 12, 9747 AD Groningen,  The Netherlands
        \and 
             Department of Physics and Astronomy, University of British Columbia, Vancouver, BC V6T 1Z1, Canada
             }

   \date{Received July 12, 2024; accepted October 20, 2024}


  \abstract
   {Cosmic voids, distinguished by their low-density environment, provide a unique opportunity to explore the interplay between the cosmic environment and the processes of galaxy formation and evolution. Data on the molecular gas has been scarce so far.}
   {In this paper we continue previous research done in the CO-CAVITY pilot project to study the molecular gas content and properties in void galaxies to search for possible differences compared to galaxies that inhabit denser structures.}
   {We observed at the IRAM 30\,m telescope the CO(1--0) and CO(2--1) emission of 106 void galaxies selected from the CAVITY survey. Together with data from the literature, we obtained a sample of 200 void galaxies with CO data. We conducted a comprehensive comparison of the specific star formation rate (sSFR = SFR/\Mstar), the molecular gas fraction (\mhtwo/\Mstar), and the star formation efficiency (SFE = SFR/$M_{\rm H_2}$) between the void galaxies and a comparison sample of galaxies in filaments and walls, selected from the xCOLD GASS survey.}
   {We found no statistically significant difference between void galaxies and the comparison sample in the molecular gas fraction as a function of stellar mass for galaxies on the star-forming main sequence (SFMS). However, for void galaxies, the SFE was found to be constant across all stellar mass bins, while there is a decreasing trend with \Mstar\, for the comparison sample. Finally, we found some indications for a smaller dynamical range in the molecular gas fraction as a function of distance to the SFMS in void galaxies.}
  {Overall, our analysis finds that the molecular gas properties of void galaxies are not very different from denser environments. The physical origin of the most significant difference that we found -- a constant SFE as a function of stellar mass in void galaxies -- is unclear and requires further investigation and higher-resolution data.
  }


\maketitle
%

\section{Introduction} 

The evolution of galaxies is, to a large extent, influenced by the formation of stars which is closely related to the availability and properties of the molecular gas \citep[][]{saintonge17, tacconi20}. This gives rise to correlations between the star formation rate (SFR), the stellar mass (\Mstar), and the molecular gas mass (\mhtwo): The so-called Kennicutt-Schmidt (KS) relation \citep{Schmidt1959, kennicutt98} is the relation between the SFR and the \mhtwo. The molecular gas main sequence (H$_2$MS) is the correlation between \mhtwo and \Mstar\ and, finally, the star-forming main sequence (SFMS) is the relation between the SFR and \Mstar\, \citep[SFMS, e.g.][]{brinchmann04,elbaz07, noeske07}. A comparison of the strengths of the individual correlations shows that the most fundamental is the KS relation, followed by the relation between molecular gas and stellar mass, whereas the SFMS seems to be the consequence of the former two correlations \citep{Baker2022}.

The distribution of a galaxy in the SFR-\Mstar\ plane gives important insights into the evolutionary pathways of galaxies. The existence of a relatively tight correlation between the SFR and the stellar mass, the SFMS, implies that the evolution of galaxies is, for a large fraction of their lives, driven by the gradual conversion of gas into stars. Phases of enhanced SFR -- starbursts (SB) -- also play a role, whereas galaxies below the SFMS have decreased or even quenched their star formation for some reason showing older stellar population ages \citep{dp2022}. Historically, galaxies in the transition zone (TZ) between the SFMS and the quiescent 'red-and-dead' galaxies well below the SFMS have been called 'green valley' galaxies because of their intermediate optical and ultraviolet colours \citep{salim07,martin07,wyder07}. However, the identification based on the distance to the SFMS is more robust as it is not affected by dust extinction and will be used in the present paper.

Apart from the molecular gas, the environment of a galaxy also has a significant influence on its evolution. The large-scale structure of the universe is characterised as a cosmic web, with walls and filaments of galaxies connecting nodes (clusters) and surrounding vast, nearly empty cosmic voids \citep{Bond1996, vdWBond2008, Cautun2014}. These voids are extensive regions (with radii from 10 to 50 h$^{-1}$ Mpc), and they constitute an integral part of the cosmic web, representing approximately 70$\%$ of the universe's volume \citep{Peebles2001, Pan12, Cautun2014, vandeWeygaert2016, Libeskind2018, Aragon2007}. Cosmic voids, defined by their pristine environment and low-density, contain galaxies that may be less affected by multiple galaxy mergers or a densely self-collapsing environment. Therefore, they are an ideal laboratory to explore the influence of the environment on the processes of galaxy formation and evolution \citep[see e.g.][]{vdW2011}. 

Observations indicate that the properties of galaxies in voids are different from galaxies in denser environments in some aspects: they tend to be bluer and of later type morphologies \citep[e.g.][]{rojas04, rojas05, park07, constantin08, Kreckel2011, Kreckel2012} and might have a slightly higher specific star formation rate (sSFR = SFR/\Mstar), than galaxies in filaments and walls \citep{beygu17}. \citet{jesus-nature} found, based on Sloan Digital Sky Survey \citep[SDSS,][] {York2000} spectra of a well-defined sample of void, wall/filament, and cluster galaxies, that the stellar mass assembly of void galaxies is slower in voids than in denser environments. \citet{dominguez23-bbb} found that on average, void galaxies have lower stellar metallicities than galaxies in denser environments. 

The content and properties of the molecular gas in void galaxies remain largely unstudied. Only a handful of studies have observed the molecular gas in void galaxies \citep{sage1997, Beygu2013, Das2016, Florez2021, dominguez22}, but most of these studies included very few (between 1 and 5) objects. \citet{dominguez22} observed 20 void galaxies from the Void Galaxy Survey \citep[VGS,][]{Kreckel2012} and this has been the largest survey so far. These studies did not find any strong evidence for differences in the molecular gas between void and filament/wall galaxies. However, the low statistical significance was a severe problem for all of them. 

To increase our knowledge about the spatially resolved properties of void galaxies, the Calar Alto Void Integral-field Treasury surveY (CAVITY\footnote{\url{https://cavity.caha.es/}}) aims to study galaxies in voids using integral field spectroscopic (IFS) data \citep{Perez2024}. The survey officially started in January 2021 at the Calar Alto Observatory (CAHA) as a legacy project, being awarded a total of 110 useful observing nights to observe $\sim$ 300 carefully selected galaxies at the 3.5 m telescope using the Potsdam Multi-Aperture Spectrophotometer (PMAS) spectrograph. 
Some first results show that void galaxies present slightly larger half-light radii (HLR), lower stellar mass surface density, and younger ages across all morphological types than galaxies in filaments and walls \citep{Conrado2024}.

In this paper we report the first results of CO-CAVITY, a comprehensive census of molecular gas via CO(1–0) and CO(2–1) spectral lines for a statistically significant sample of 106 galaxies selected from the CAVITY sample and conducted as part of the extended project (CAVITY+), by using the 30m radio telescope operated by the Institut de Radioastronomie Millim\'etrique (IRAM). Additionally, an appropriate comparison sample of filament/wall galaxies has been selected in order to quantify possible differences between void galaxies and those in denser environments.

The structure of this paper is as follows. We describe the void galaxy sample and the comparison sample in Sect.~\ref{sec:samples}. The observations and the data reduction are described in Sect.~\ref{sec:data}. The results and discussion are presented in Sect.~\ref{sec:results} and Sect.~\ref{sec:discussion}, respectively. We outline our main conclusions in Sect.~\ref{sec:conclusions}. In Appendix~\ref{app:co10_spectra} we present the CO line emission spectra of the observed
galaxies, and, in Appendix~\ref{app:SFR} we compare different SFR tracers for galaxies. To compute distances, we adopt a flat $\Lambda$CDM cosmology, with $\Omega_{\rm m}$\,=\,0.30, $\Omega_\Lambda$\,=\,0.70, and the Hubble constant \makebox{H$_0$\,=\,70 \kms Mpc$^{-1}$}.

\section{Samples}
\label{sec:samples}

\subsection{IRAM 30\,m sample (CO-CAVITY)}
\label{subsec:iram-sample}

We selected the targets for the IRAM 30\,m CO-CAVITY survey from those galaxies in CAVITY \citep{Perez2024} that already had IFU data at the time of the IRAM 30\,m observations, either from  CAHA (86 objects) and/or from MaNGA \citep[Mapping Nearby Galaxies at Apache Point Observatory][]{Bundy2015} (20 objects). 

The selection criteria are described and justified in detail in \citet{Perez2024}. Here, we give a short summary. The CAVITY sample is selected from the \citet{Pan12} catalogue of void galaxies based on SDSS Data Release 7 data (SDSS DR7). The voids were selected to be in a redshift range between $0.005$ and  $0.05$. A final sample of 15 voids was chosen, carefully selected to represent the full range of void parameters, and to contain at least 20 galaxies that fit the galaxy selection criteria. As for the galaxies in the voids, they were restricted to intermediate inclinations. To ensure that the final selection of galaxies represents indeed objects belonging to voids, firstly galaxies listed in the cluster catalogue of \citet{tempel17} were excluded and, secondly, galaxies were limited to the (relatively) inner regions of voids. For this, the effective radius parameter, $R_{\rm eff}$ was used which corresponds to the radius of a spherical void of equal volume. Only galaxies with $R_{\rm eff} < 0.8$ were included. 
 
The galaxies in CAVITY span stellar masses between log~\Mstar/\Msun = 8 -- 11 and SFRs between \makebox{SFR = 10$^{-2}$ -- 10$^{0.7}$ \Msun\,yr$^{-1}$}. Out of this sample, we selected objects on the SFMS \citep[defined as in Eq. (1) in ][]{janowiecki20} and in the transition zone (TZ) below the SFMS. We excluded quiescent galaxies lying more than 1 dex below the SFMS, in order to restrict our analysis to at least moderate actively star-forming galaxies. 
Furthermore, we excluded objects with low stellar masses ($<$ 10$^9$\,\Msun) because this mass bin is only sparsely populated by the CAVITY observed sample and galaxies tend to have low metallicities (12+log~O/H $<$ 8.4) for which the Galactic CO-to-molecular gas mass conversion factor might not apply \citep{bolatto13}. In total, we observed 106 CAVITY galaxies. The main properties of the IRAM 30\,m CO-CAVITY sample are summarised in Table~\ref{tab:properties}.

\begin{table}
\caption{\label{tab:properties} Properties for the IRAM-30\,m CO-CAVITY galaxies.}
\begin{adjustbox}{max width=0.5\textwidth}
\begin{tabular}{cccccc}
\noalign{\smallskip} \hline \noalign{\medskip}
CAVITY ID &  RA (J2000.0) &  DEC (J2000.0) & z$_{SDSS}$  &  D$_L$  & R$_{90,r}$\\
          &   (hh:mm:ss)  &  (dd:mm:ss.s)  &             & [Mpc] & [arcsec] \\      
(1)       &  (2)          & (3)            & (4)         & (5)   & (6) \\
\noalign{\smallskip} \hline \noalign{\medskip}
3427 & 8:44:03.80 & 50:11:08.4 & 0.0425 & 187.71 &  7.6  \\ 
3670 & 8:44:27.74 & 51:26:49.8 & 0.0449 & 198.92 &  8.0  \\ 
7926 & 9:05:48.41 & 51:06:13.1 & 0.0455 & 201.64 & 10.2  \\ 
7948 & 9:04:02.48 & 51:28:13.8 & 0.0409 & 180.75 & 19.3  \\ 
8556 & 8:59:00.16 & 49:10:23.1 & 0.0406 & 179.40 & 11.3  \\ 
8595 & 9:00:55.58 & 50:14:07.5 & 0.0412 & 182.14 &  7.7  \\ 
8645 & 9:05:28.85 & 51:24:49.2 & 0.0412 & 182.08 &  5.9  \\ 
8646 & 9:05:24.32 & 51:27:36.8 & 0.0406 & 179.48 &  8.2  \\ 
8728 & 8:54:34.17 & 50:45:56.4 & 0.0444 & 196.59 & 13.8  \\ 
8996 & 8:35:03.66 & 47:56:09.9 & 0.0432 & 191.15 &  9.7  \\ 
...  & ...        & ...        & ...     & ... & ... \\
\noalign{\smallskip} \hline \noalign{\medskip}
\end{tabular}
\end{adjustbox}
\tablefoot{(1) Catalogue ID in the CAVITY sample. (2) Right Ascension (RA). (3) Declination (DEC). (4) Redshift (from SDSS DR7). (5) Luminosity distance in Mpc. (6) The r-Petrosian radius R$_{90,r}$ (from SDSS DR7) which is the radius enclosing 90 percent of the Petrosian flux in the $r$ band from SDSS. The full table is available online at the CDS.}
\end{table}

\subsection{CO-void galaxy sample (CO-VG)}

The final void galaxy sample was constructed as follows: in addition to the 106 CAVITY galaxies, we included the sample from \citet{dominguez22}, selected from the Void Galaxy sample \citep[VGS,][]{Kreckel2012}, which contains 20 objects observed in CO(1-0) and CO(2-1) with the IRAM 30\,m telescope. Four galaxies with stellar masses less than $10^9$ \Msun\ were excluded from this subset to maintain consistency with  CO-CAVITY and the comparison sample. Consequently, 16 VGS galaxies were included, amounting to 122 void galaxies. Among the new objects, six galaxies, identified as VGS32, VGS49, VGS50, VGS53, VGS56, and VGS57 in \citet{dominguez22}, belong to the CAVITY parent sample. Therefore, we use their corresponding CAVITY identification numbers: CAVITY 23926, 43490, 42665, 37424, 43294, and 34101, respectively.

Additionally, in order to improve the statistics, we have included a subset of 89 galaxies from the xCOLD GASS sample \citep{saintonge17} located in voids as revealed by cross-matching with the catalogue of \citet{Pan12}. Before including them, we applied the same selection criterion as for the CAVITY sample, and removed galaxies whose distance to the centre of the void, characterised by the effective radius parameter (R$_{\rm eff}$), is larger than 0.8 \citep{Perez2024}. 

Finally, we excluded  11 void galaxies classified as AGN in the BPT diagram (see Sect. \ref{subsec:metal-agn}; 4 from the IRAM 30\,m CO-CAVITY survey and 7 from the xCOLD GASS subset of void galaxies sample, resulting in a total of 200 void galaxies for more details). From here on, we will refer to this final sample as the CO-void galaxy sample (CO-VG).

\subsection{CO comparison sample (CO-CS)}

We selected the xCOLD GASS sample as our comparison sample. The xCOLD GASS survey is a mass and
redshift selected (\Mstar $> 10^9$ \Msun, and \makebox{$0.01 < z < 0.05$)} 
IRAM 30\,m telescope CO legacy survey consisting of 532 nearby galaxies. This survey does not have any specific environmental selection criteria and therefore contains galaxies in voids, filaments, walls and clusters. Since we want to compare the CO-VG sample only to galaxies in filaments and walls, we remove objects that are located in voids \citep{Pan12} and in clusters 
\citep[those with more than 30 objects i.e. N\,$>$\,30,][]{tempel17}, resulting in a sample of 353 galaxies. We exclude cluster galaxies in order to have a comparison sample with a clearly defined environment (filaments and walls) and because clusters represent an extreme environment that can affect the molecular gas content as shown in previous studies \citep[e.g.][]{zabel2019}. Additionally, in order to constrain the sample to star-forming galaxies, we exclude 88 objects classified as AGN in the BPT diagram (see Sect.~\ref{subsec:metal-agn}). The final CO-CS is composed of 265 galaxies.

\section{Data}
\label{sec:data}

\subsection{CO observations with the IRAM 30\,m telescope and data reduction}
\label{sec:coobs}

We have carried out a CO spectroscopic survey of a sample of 106 CAVITY galaxies
with the IRAM 30\,m telescope at Pico Veleta (Spain) to observe the $^{12}$CO(1--0) and in parallel 
$^{12}$CO(2--1) line emission. The corresponding rest frequencies are 115.271\,GHz, 
and 230.538\,GHz for CO(1--0) and CO(2--1) respectively. The observations were carried out between 
December 2021 and February 2023, within the 
projects 170-21, 083-22 and 164-22. We used the E090 and the E230 EMIR bands simultaneously in combination 
with the Fourier Transform Spectrometer (FTS) at a frequency resolution of 0.195 MHz 
(corresponding to a velocity resolution of $\sim$ 0.5 \kms\ for CO(1--0) at the sky frequency of our observations) 
and with the autocorrelator WILMA with a frequency resolution of  2~MHz (corresponding to a velocity resolution 
of  $\sim$ 5 \kms\ for CO(1--0)). The broad bandwidth of the receiver (2 $\times$ 16 GHz) and backends (8 $\times$ 4 GHz for FTS and 4 $\times$ 4 GHz for WILMA) allow the observations to be grouped into similar 
galaxy redshifts. The observed sky frequencies, i.e. the line frequencies after taking into account the redshift of the objects, ranged between 
108.2~GHz and 114.4~GHz for the CO(1--0) line and 219.7~GHz and 231.5~GHz for the CO(2--1) line. The 
observations were done in wobbler switching mode with a wobbler throw of 60\arcsec\ in azimuthal direction. We checked for each galaxy that the off-position was well outside the galaxy. Half Power Beam Width (HPBW) are $\sim$ 22$\arcsec$ and $\sim$ 11$\arcsec$ for CO(1--0) and for CO(2--1), respectively.

Each object was observed until it was detected with a signal-to-noise ratio (S/N) of at least 5 or until a root-mean-square 
noise (rms) of $\sim 1.5 $ mK (T$_{\rm mb}$) was achieved for the CO(1--0) line at a velocity resolution 
of 20 km s$^{-1}$. Only five galaxies (CAVITY 30216, 33891, 39573, 41296, 46746) undetected in CO(1--0) have a slightly higher rms (from 1.6 to 2.2 mK) than this limit. The on-source integration times per object ranged between 30~minutes and 4 hours. Pointing was monitored on nearby quasars every 60 -- 90  minutes. During the observation period, the weather conditions were in general good; for periods where the weather conditions were variable, we selected observations with a pointing accuracy better than 4--5\arcsec. Data taken in poorer conditions (i.e., with worse pointing accuracy or presenting corrupted baselines) were rejected.
The temperature scale used is the main beam temperature $T_{\rm mb}$, which is defined as 
$T_{\rm mb} = (F_{\rm eff}/B_{\rm eff})\times T_{\rm A}^*$. For the observed frequencies for CO(1--0) and for CO(2--1) the IRAM forward efficiency, $F_{\rm eff}$, is 0.95 and 0.91 respectively, and the beam efficiency, $B_{\rm eff}$, is 0.79 and 0.56, respectively\footnote{\url{https://publicwiki.iram.es/Iram30mEfficiencies}}. The mean system temperature for the observations was 170~K for CO(1--0) and 448~K for CO(2--1) on the $T_{\rm A}^*$ scale. 

The data were reduced in the standard way via the CLASS software in the GILDAS package\footnote{\url{http://www.iram.fr/IRAMFR/GILDAS}}. We first discarded poor scans and data taken in poor weather conditions (e.g. with large pointing uncertainties) and then subtracted a constant or linear baseline from individual integrations. Some observations taken with the FTS backend were affected by platforming, that is, the baseline level changes abruptly at one or two positions along the band. This effect could be reliably corrected because the baselines in between these (clearly visible) jumps were 
linear and could be subtracted from the different parts individually, using the 
{\tt FtsPlatformingCorrection5.class} procedure, developed by IRAM. We then averaged the spectra and box-smoothed them to a resolution of 20~\kms.

We present the detected or tentatively detected spectra in Appendix~\ref{app:co10_spectra}. For each spectrum, the zero-level line width was determined visually from the final spectrum, by measuring the total width of those channels that are above the zero-level baseline.
The velocity-integrated line intensities, \icoone\ and \icotwo,  were calculated by summing the individual channels in between these limits (the colour-shaded channels in Figs.~\ref{co10-emission} and \ref{co21-emission}. The error was calculated as:

\begin{equation}
{\rm err}(I_{\rm CO}) = {\rm rms} \times \sqrt{\delta \rm{V} \ \Delta V},
\end{equation}

\noindent where $\delta \rm{V}$ is the channel width (in km s$^{-1}$), $\Delta$V the zero-level line 
width (in km s$^{-1}$), and rms the root mean square noise (in K) which was determined from the the line-free channels in a range between 1000 km s$^{-1}$ on either side of the CO line. 

To test the robustness of the measurement of the \icoone, we also fitted the spectra by a single-Gaussian profile. This yielded a very good linear relation between  \icoone\ derived from both methods with a mean value of  I$_{\rm CO, gaussian}$/I$_{\rm CO, sum-channels} = 1.0\pm 0.2$.

For non-detections we set an upper limit as:

\begin{equation}
I_{\rm CO} < 3 \times {\rm rms} \times \sqrt{\delta \rm{V} \ \Delta V}.
\end{equation}

For the non-detections, we  assumed a line width of $\Delta$V = 300 \kms\ which is close to the mean velocity width found for CO(1--0) and CO(2--1) in the sample (mean $\Delta \rm{V}$ = 340 \kms\ with a standard deviation of $\sim$ 106\,\kms). We considered spectra with a S/N ratio (defined as the ratio between the velocity integrated intensity of the line and its error, I$_{\rm CO}$/err(I$_{\rm CO}$) of S/N $> 5$ as firm detections and those with a S/N ratio in the range of 3--5 as tentative detections. In the statistical analysis, we count tentative detection as detections.  The results of our CO(1--0) observations are listed in Table~\ref{tab:ico}. Regarding CO(1--0), we obtain 64 detections, 9 tentative detections, and 33 non-detections, and for CO(2--1) we obtain 63 detections, 8 tentative detections, and 35 non-detections. There are seven galaxies (CAVITY 16769, 26668, 37605, 39573, 40774, 51442, and 65887) where CO(1--0) emission was undetected but CO(2--1) emission was tentative or firmly detected. The reason for this could be due to the less severe beam dilution for I$_{\rm CO(2-1)}$, since the beam at 1\,mm is smaller. In addition to the statistical error of the velocity-integrated line intensities, a calibration error of 15\% for CO(1--0) and 30\% for CO(2--1) was taken into account \citep[see][]{lisenfeld19}. 

\subsubsection{CO data quality control}
 
After generating the final CO(1--0) and CO(2--1) emission spectra, we conducted a quality control analysis. This analysis involved inspecting each CO spectral profile and searching for possible errors and inconsistencies. In particular, we checked the following points: (i) we visually inspected the spectra to make sure that the detections are convincing with no indications of artefacts; (ii) we confirmed that the baselines are flat, as wavy baselines can indicate an incorrect baseline subtraction or contamination with poor quality data; (iii) we checked that the defined velocity window, accurately constraining the entire CO emission, was consistent with the recession velocity derived from the optical redshift; (iv) we checked the consistency between the CO(1--0) and CO(2--1) lines: The CO(1--0) line widths are expected to be similar to or larger than the CO(2--1) line widths due to the smaller beam at the higher frequency of CO(2--1), which is emitted from a smaller area.

As a part of this analysis, we produced plots of the CO(1--0) and CO(2--1) lines at various velocity resolutions (10, 20, and 40~\kms) to gain a different perspective of the data increasing the S/N ratio. As mentioned before, one of the requirements was to observe each object until it was detected with a S/N ratio of at least 5 or until a rms of 1.5 mK (T$_{mb}$) was achieved for the CO(1-0) line at a velocity resolution of 20 km s$^{-1}$. A priori, this requirement ensured high quality for the CO data in general, preventing the rejection of any observed galaxy.

\begin{table*}
\centering
\caption{\label{tab:ico} CO emission line intensities for the IRAM-30\,m CO-CAVITY.}
\begin{tabular}{ccccccccc}
\noalign{\smallskip} \hline \noalign{\medskip}
CAVITY ID &  I$_{\rm CO(1-0)}$   &  rms & S/N &  $\Delta V_{\rm CO(1-0)}$ &  I$_{\rm CO(2-1)}$   &  rms & S/N  &  $\Delta V_{\rm CO(2-1)}$  \\
          &  [K km s$^{-1}$] & [mK] &         &    [km s$^{-1}$]      &  [K km s$^{-1}$] & [mK]     &      &    [km s$^{-1}$] \\
(1)       &  (2)             & (3)  & (4)     & (5)                   & (6)               & (7)     &  (8) & (9) \\
\noalign{\smallskip} \hline \noalign{\medskip}
3427 & 1.98 $\pm$ 0.21 & 2.24 & 9.6 & 403 & 5.41 $\pm$ 0.35 & 3.48 & 15.3 & 486\\ 
3670 & 0.56 $\pm$ 0.11  & 1.25 & 5.0 & 381 & $<$ 0.68 & 2.86 & -- & 300 \\ 
7926 & 3.28 $\pm$ 0.16 & 1.48 & 20.7 & 547 & 4.88 $\pm$ 0.25 & 2.43 & 19.9 & 484\\ 
7948 & 0.92 $\pm$ 0.14 & 1.72 & 6.4 & 331 & 1.49 $\pm$ 0.16 & 2.00 & 9.4 & 296\\ 
8556 & $<$ 0.30  & 1.28 & -- & 300 & $<$ 0.40 & 1.66 & -- & 300 \\ 
8595 & $<$ 0.34  & 1.44 & -- & 300 & $<$ 0.56 & 2.36 & -- & 300 \\ 
8645 & 3.69 $\pm$ 0.27 & 3.35 & 13.8 & 303 & 11.62 $\pm$ 0.59 & 7.00 & 19.7 & 336\\ 
8646 & 1.66 $\pm$ 0.10 & 1.09 & 17.4 & 365 & 4.28 $\pm$ 0.18 & 1.87 & 24.3 & 419\\ 
8728 & 0.71 $\pm$ 0.09 & 1.22 & 7.6 & 279 & 0.82 $\pm$ 0.17 & 2.53 & 4.7 & 221\\ 
8996 & $<$ 0.33  & 1.36 & -- & 300 & $<$ 1.02 & 4.27 & -- & 300 \\ 
...  & ...       & ...  & ...& ... & ...   &  ...& ...&  ... \\
\noalign{\smallskip} \hline \noalign{\medskip}
\end{tabular}
\tablefoot{(1) Catalogue ID in the CAVITY sample. (2) and (6) Velocity-integrated intensity and its statistical error for the CO(1--0) and CO(2--1) emission, respectively. (3) and (7) Root-mean-square noise at a velocity resolution of $\Delta$V  = 20 km s$^{-1}$. (4) and (8) Signal-to-noise ratio (S/N) for CO(1--0) and CO(2--1) emission, respectively. (5) and (9) Zero-level line widths for CO(1--0) and CO(2--1) lines. The full table is available online at the CDS.}
\end{table*}

\subsubsection{Aperture correction}
\label{sec:molecular_gas_mass}

Each galaxy was observed with a single pointing at the centre. The IRAM CO(1--0) beam of 22\arcsec\ is, in general, sufficient to cover a significant fraction of the galaxy (see Fig.~\ref{sample-histo} bottom). However, the fraction covered by the beam varies for each galaxy depending on its apparent size. Therefore, it is necessary to apply a correction for the emission outside the beam. We calculated this aperture correction following the procedure outlined in \citet{lisenfeld11}, assuming an exponential distribution of the CO flux:

\begin{equation}
S_{\rm CO}(r) = S_{\rm CO,center}\propto \exp(-r/r_{\rm e}),
\label{eq:Ico_r}
\end{equation}

\noindent where $S_{\rm CO,center}$ is the CO(1--0) flux in the central position derived from the measured \icoone\ applying the 
$T_{\rm mb}$-to-flux conversion factor of the IRAM 30\,m telescope (5 Jy/K). 
\citet{lisenfeld11} adopted an exponential scale length of \makebox{$r_{\rm e} = 0.2\,\times r_{\rm 25}$},
where \riso\ is the major optical isophotal radius at 25 mag arcsec$^{-2}$.  We use $r-$Petrosian radius R$_{90,r}$ from the SDSS and the relation $r_{\rm 25}\,=\,1.7\,\times\,R_{90,r}$ as previously calculated in \citet{dominguez22}. The resulting aperture correction factors, \faper, defined as the ratio between $S_{\rm CO,centre}$ and the total aperture-corrected flux $S_{\rm CO,tot}$, lie between 1.02 and 1.57. The values of \faper\ are listed in Table~\ref{tab:MH2mass}.

\subsubsection{Molecular gas mass and \alphaco} 
\label{moleculargas}

We calculate the molecular gas mass from the CO(1--0) luminosity, \lco, following \citet{solomon97} as:

\begin{equation}
L^\prime_{\rm CO} [{\rm K \, km\, s^{-1} pc^{2}}]= 3.25 \times 10^7\, S_{\rm CO,tot} \nu_{\rm rest}^{-2} D_{\rm L }^{2} (1+z)^{-1},
\label{eq:lco}
\end{equation}
where  $S_{\rm CO, tot}$ is the aperture-corrected CO(1--0) line flux (in Jy\,\kms), 
$D_{\rm L }$ is the luminosity distance in Mpc, $z$ the redshift,
and $\nu_{\rm rest}$ is the rest frequency of the line in GHz.
We then calculate the molecular gas mass, M$_{\rm H_2}$ as:

\begin{equation}
M_{\rm H_2} [M_\odot]= \alpha_{\rm CO} L^\prime_{\rm CO}.
\label{eq:mmol}
\end{equation}

As previously mentioned, there are 7 galaxies (CAVITY 16769, 26668, 37605, 39573, 40774, 51442, and 65887) undetected in CO(1--0), but detected in CO(2--1). For these galaxies, we estimate $I_{\rm CO(1-0)}$ 
from $I_{\rm CO(2-1)}$ by deriving the theoretically expected value of the observed ratio \makebox{$R_{\rm 2-1,theo} = I_{\rm CO(2-1)}/ I_{\rm CO(1-0)}$} following the procedure of \citet[][their Appendix A] {dominguez22}. As explained in detail in Sect.~\ref{sec:line-ratio}, this ratio depends on the intrinsic central brightness ratio that we adopt as $\bar{T}_{Bc2-1} / \bar{T}_{Bc1-0}$ = 0.8 \citep{leroy09}, and on the respective beam and source sizes. Taking the corresponding values of $r_{\rm e}$ and 
$i$ (i.e. disk galaxy inclination angle) for each galaxy,
we derive $R_{\rm 2-1,theo}$ of 2.2, 2.3, 2.0, 2.2, 2.4, 2.5, and 2.4 for CAVITY 16769, 26668, 37605, 39573, 40774, 51442, and 65887, respectively, yielding a predicted I$_{\rm CO(1-0)}$ of 0.80, 0.70, 0.62, 1.18, 0.83, 0.69, and 0.84, respectively.

 The conversion factor \alphaco\ is known to vary as a function of metallicity. \cite{accurso17} studied integrated scaling relations involving [CII] 158~$\mu m$ data from {\it Herschel} PACS and CO(1-0) for a subsample of xCOLD GASS combined with galaxies from the Dwarf Galaxy Survey \citep{madden13, cormier14}. They showed that \alphaco\ varies as a function of two parameters: i) the metallicity, derived from the SDSS central fiber spectra, and, to a minor extent, ii) the distance to the star-forming main sequence (see Eq.~\ref{eq:dist_SFMS} for its definition). This prescription was used to calculate the molecular gas mass in the xCOLD GASS sample. Thus, in order to be consistent with our comparison sample, we also use a variable \alphaco\ factor (which takes the Helium fraction into account) following the \citet[][Eq. 25]{accurso17} prescription. For this, we adopt the SFMS derived by \cite{janowiecki20} for the xCOLD GASS sample and calculate the gas-phase metallicity of the galaxies using the calibration from \cite{pp04} (more details in \ref{subsec:metal-agn}). 
 
 The CO luminosity, the conversion factor, and the molecular gas masses, calculated with this variable \alphaco\ factor are listed in Table \ref{tab:MH2mass}. 

\begin{table*}
\centering
\caption{\label{tab:MH2mass} L'$_{\rm CO10, obs}$ luminosity, $\alpha_{\rm co}$, molecular gas mass and aperture correction factor for the CO-VG sample.} 
\begin{tabular}{ccccc}
\noalign{\smallskip} \hline \noalign{\medskip}
CAVITY &  L'$_{\rm CO10, obs}$             & $\alpha_{\rm co}$ & log M$_{\rm H_2}$ & \faper\ \\
 ID &  [10$^8$ K km s$^{-1}$ pc $^{2}$]   & [\Msun (K km s$^-1$ pc$^2$)$^{-1}$]    &   [\Msun]   &    \\
 (1)      &    (2)        &   (3)        &  (4)              & (5) \\
\noalign{\smallskip} \hline \noalign{\medskip}
3427 &  8.18 $\pm$  1.04 & 3.1 &  9.43 $\pm$  0.05 & 1.05 \\ 
3670 &  2.60 $\pm$  0.58 & 3.5 &  8.98 $\pm$  0.10 & 1.05 \\ 
7926 & 15.58 $\pm$  1.11 & 4.1 &  9.84 $\pm$  0.03 & 1.09 \\ 
7948 &  3.52 $\pm$  0.63 & 2.8 &  9.07 $\pm$  0.08 & 1.18 \\ 
8556 & $<$  1.12 & 3.1 & $<$  8.57 & 1.06  \\ 
8595 & $<$  1.30 & 3.8 & $<$  8.72 & 1.05  \\ 
8645 & 14.36 $\pm$  1.36 & 3.5 &  9.71 $\pm$  0.04 & 1.03 \\ 
8646 &  6.28 $\pm$  0.50 & 3.1 &  9.31 $\pm$  0.03 & 1.06 \\ 
8728 &  3.20 $\pm$  0.49 & 3.3 &  9.09 $\pm$  0.07 & 1.17 \\ 
8996 & $<$  1.35 & 3.0 & $<$  8.65 & 1.08  \\ 
...  &  ...      & ...   & ... & ... \\
\noalign{\smallskip} \hline \noalign{\medskip}
\end{tabular}
\tablefoot{(1) Catalogue ID in the CAVITY sample. (2) Beam-integrated CO (1–-0) line luminosity. (3) CO-to-H$_2$ conversion factor, calculated using the \citet[][Eq. 25]{accurso17} prescription. (4) Molecular gas mass and total error. (5) Aperture correction factor. The molecular gas mass has been derived from the CO(1--0) line emission except for those galaxies marked with $^*$ where the CO(1--0) was not detected and the molecular gas mass was derived from the CO(2--1) line emission. The full table is available online at the CDS.}
\end{table*}

\subsection{Additional data}

\subsubsection{SDSS optical data} 
\label{subsec:metal-agn}

We use SDSS optical spectroscopic data \citep[SDSS-DR7][]{Abazajian2009} for three purposes: (i) to obtain the oxygen abundance as an indicator for the gas metallicity of a galaxy, (ii) to derive the SFR based on H$\alpha$ as described in Sect. \ref{subsec:SFR-calculation}, and (iii) to identify galaxies dominated by AGN or shocks according to their optical emission lines which were excluded from the statistical study because the SFR and stellar mass determination might be erroneous due to the influence of the AGN on the UV, IR, and line emission.

We estimate the oxygen abundance from the [OIII]$\lambda 5007 \AA$/H$\beta$ and the [NII]$\lambda 6583 \AA$/H$\alpha$ emission-line ratios using the prescription from \cite{pp04}: 

\begin{equation}
12+\log {\rm O/H} = 8.73 - 0.32 \times O3N2,
\label{eq:metal}
\end{equation}
(with O3N2 = ([OIII]$\lambda 5007 \AA$/H$\beta$)/ ([NII]$\lambda 6583 \AA$/H$\alpha$)) to calculate the conversion factor \alphaco\ (see Sect. \ref{moleculargas}) for the CO-VG sample following the same methodology as for the XCOLD GASS sample \citep{accurso17, saintonge17}. On this scale, a value of 12 + log~O/H = 8.69 corresponds to solar metallicity \citep{Asplund2004}. 

\begin{figure*}
\sidecaption
\centerline{
\includegraphics[width=0.5\textwidth]{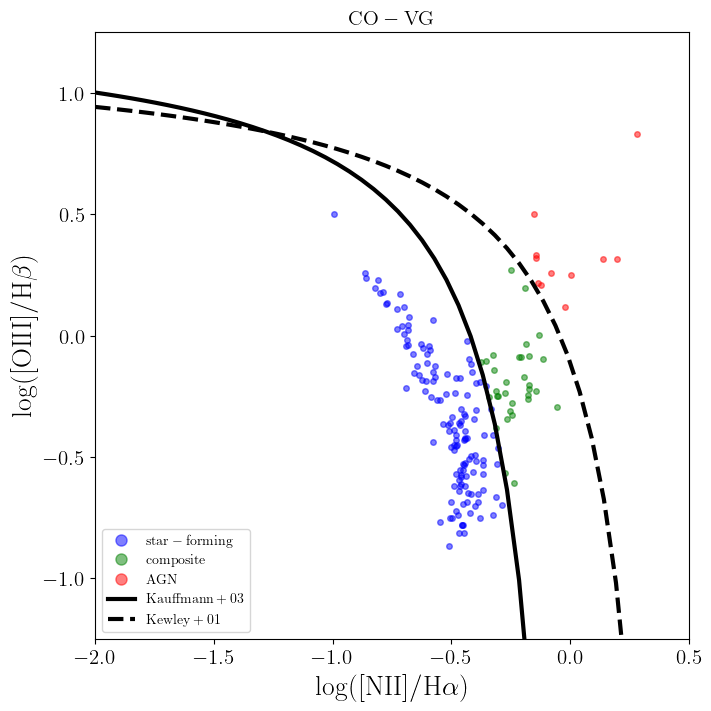}
\includegraphics[width=0.5\textwidth]{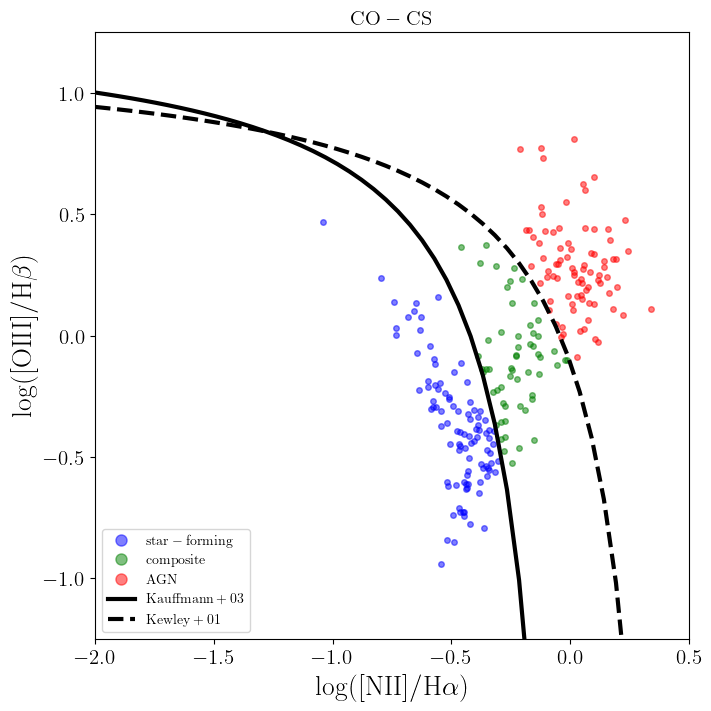}
}
\caption{BPT diagnostic diagram for CO-VG galaxies (left) and CO-CS galaxies (right). Blue points represent star-forming galaxies, green points represent composite galaxies and red points represent AGN galaxies. The dashed line represents the demarcation proposed by \cite{Kewley01}, while the solid line corresponds to the curve introduced by \cite{Kauffmann03a}.}
\label{BPT}
\end{figure*}

Additionally, we use a BPT diagram ([OIII]5007/H${\beta}$ versus NII6583/H${\alpha}$, \citealp{baldwin81}) as a diagnostic tool for classifying the star-forming, composite, and AGN galaxies. 
Figure ~\ref{BPT} displays the BPT diagram for both the CO-VG sample and the comparison sample. We exclude those galaxies that are classified as AGNs based on the \cite{Kewley01} definition in the BPT diagram, which removes 88 galaxies from the comparison sample and 11 galaxies from the CO-VG sample. Transition-type galaxies are included. Whereas a single diagnostic diagram is not enough to reliably identify AGN galaxies, this diagram reliably identifies galaxies dominated by SF \citep{baldwin81, Kewley01, Kauffmann03a, brinchmann04, Gunawardhana2013} which is the restriction that we need to apply to our analysis.

\subsubsection{Stellar masses (\Mstar)}

The stellar masses for both the comparison sample \citep{saintonge17} and the CO-VG sample are obtained from the MPA/JHU catalog, a comprehensive dataset provided by the Max Planck Institute for Astrophysics (MPA) in collaboration with Johns Hopkins University (JHU) {\footnote{\url{http://home.strw.leidenuniv.nl/~jarle/SDSS/}}} \citep{Kauffmann03b, Brinchmann2004, Tremonti2004, salim07}. The determination of total stellar mass is done through fits to the SDSS five-band ($ugriz$) photometry based on \cite{Kauffmann03b}.

\subsubsection{Star formation rate (SFR)}
\label{subsec:SFR-calculation}

For the CO-VG sample, we calculated the SFR using the method of \cite{dp17}, based on H${\alpha}$ and H${\beta}$ fluxes from the SDSS data because the SDSS fibre has a limited size of 2.5". The H${\alpha}$ fluxes were corrected using an empirical aperture correction derived from integral-field spectroscopic (IFS) observations of nearby galaxies \citep{Iglesias-Paramo2016, Iglesias-Paramo2013} and extinction was corrected using the H${\alpha}$/H${\beta}$ ratio. The method is similar to that of MPIA/JHU \citep{brinchmann04}, but it incorporates an improved aperture correction, resulting in better agreement with other methods for low-mass galaxies (see Appendix~\ref{app:SFR}).

This method differs from the determination of the SFR for the comparison sample, which is based on the near-ultraviolet (NUV) flux from the Galaxy Evolution Explorer \citep[GALEX,][]{Martin05, Morrissey2007}
and the mid-infrared (MIR) emission from the Wide-field Infrared Survey Explorer \citep[WISE,][]{Wright10}, in a SFR "ladder" approach as described in \cite{janowiecki17}. For consistency with previous studies involving xCOLD GASS, we are going to use this value of the SFR  for the comparison sample.  

We cannot apply the same method to the CO-VG sample because of the lack of GALEX or WISE data for a considerable number of objects (49 galaxies). 
We therefore chose to use the H${\alpha}$ fluxes as a SFR tracer for CO-VG, following \citet{dp17}. However, also for this method, there were a few galaxies (27 galaxies; 8 galaxies from the IRAM 30\,m CO-CAVITY survey and 19 from the xCOLD GASS void galaxies subset) without the necessary high-quality H${\alpha}$/H${\beta}$ lines. For those cases, we used GALEX and WISE data (for 24 galaxies) to determine the SFR following the procedure described in \cite{janowiecki17}, and for the remaining 3 objects without NUV and/or W4 data we calculated the SFR from the W3 data, following the prescription described in \cite{leroy19}. 

A comparison between these methods for galaxies with available data is given in Appendix~\ref{app:SFR}, obtaining a satisfactory agreement. The mean difference between the values of SFR$_{\rm NUV+MIR}$ and SFR$_{\rm H\alpha}$ is 0.01 dex, and there is a difference of 0.1 dex between the values of SFR$_{\rm MIR}$ and SFR$_{\rm H\alpha}$. 

\begin{table}
\caption{\label{tab:properties2} Properties for the CO-VG sample.}
\begin{adjustbox}{max width=0.5\textwidth}
\begin{tabular}{cccc}
\noalign{\smallskip} \hline \noalign{\medskip}
CAVITY ID &  log~\Mstar &  log~SFR & 12 + log~O/H \\
          & [\Msun]          &  [\Msun yr$^{-1}$] &           \\
(1)       & (2)              & (3)                &(4)         \\
\noalign{\smallskip} \hline \noalign{\medskip}
3427 & 10.52 $\pm$  0.10 &  0.59 $\pm$  0.01 & 8.80 \\ 
3670 & 10.27 $\pm$  0.09 & -0.32 $\pm$  0.01 & 8.74 \\ 
7926 & 10.72 $\pm$  0.10 &  0.54 $\pm$  0.01 & 8.72 \\ 
7948 & 10.59 $\pm$  0.09 & -0.08 $\pm$  0.01 & 8.81 \\ 
8556 & 10.34 $\pm$  0.09 & -0.80 $\pm$  0.16 & 8.76 \\ 
8595 & 10.12 $\pm$  0.08 & -0.05 $\pm$  0.01 & 8.73 \\ 
8645 & 10.45 $\pm$  0.11 &  1.31 $\pm$  0.01 & 8.84 \\ 
8646 & 10.32 $\pm$  0.09 &  0.34 $\pm$  0.01 & 8.82 \\ 
8728 &  9.96 $\pm$  0.08 &  0.02 $\pm$  0.01 & 8.78 \\ 
8996 &  9.96 $\pm$  0.09 & -0.48 $\pm$  0.14 & 8.78 \\ 
 ... &   ...             & ...               & ... \\           
\noalign{\smallskip} \hline \noalign{\medskip}
\end{tabular}
\end{adjustbox}
\tablefoot{(1) Catalogue ID in the CAVITY sample. (2) Stellar mass from MPA-JHU catalog \citep{Kauffmann03b, Brinchmann2004, Tremonti2004, salim07}. (3) H${\alpha}$ based SFR calculated following the method presented in \cite{dp17}. (4) Gas-phase metallicity 12 + log O/H using both the [N II]/H${\alpha}$ and [O III]/H${\beta}$ line ratios and the calibration of \cite{pp04}. The full table is available online at the CDS.}
\end{table}

\section{Results}
\label{sec:results}

In this section, we examine different properties of void galaxies in comparison with the comparison sample CO-CS, such as the specific star-formation rate (sSFR = SFR/\Mstar), the molecular gas mass fraction (\mhtwo/\Mstar), and the star formation efficiency (SFE = SFR/\mhtwo), as a function of the stellar mass of the galaxy \Mstar. To further refine this comparison, we divide the sample into four different categories based on their position relative to the SFMS (see Sect.~\ref{sec:SFMS} for details on the definition of the SFMS): Starbust (SB) galaxies that lie 0.3 dex or more above the SFMS, star-forming (SF) galaxies within the SFMS ($\pm$ 0.3 dex), transition-zone (TZ) galaxies down to SFMS - 0.155 dex, and red-sequence (RS) galaxies lying below this limit \citep[see][]{janowiecki20}.

Additionally, we have subdivided the sample into 5 different mass bins, $\log$\,\Mstar\,[{\it M$_{\odot}$}]\,=\,9.0--9.5, 9.5--10.0, 10.0--10.5, 10.5--11.0, and 11.0--11.5. In order to obtain a robust statistical significance, we employ the Kaplan-Meier estimator \citep{KM58}, which takes into account the upper limits when determining the mean value. As an additional test, the Kolmogorov–Smirnov (KS) test was also applied to each stellar bin to judge the significance on any differences between the samples. Here, upper limits were treated as detections. \footnote{The KS test is a statistical tool used to compare the characteristics of two distinct sample distributions. The KS test evaluates whether two samples come from the same mother sample. A p-value below 0.05 indicates with reliability higher than 95\% that both samples come from different mother samples, whereas for higher p-values, no firm conclusions can be drawn.}

Despite variations in the distribution of the samples (see Fig.~\ref{sample-histo}), the differences are minimal when considering only galaxies within the SFMS, except for bins with mass log \Mstar/\Msun > 10.5, where the CO-CS sample exhibits a larger population of galaxies in comparison to the CO-VG sample. Therefore, for a straightforward comparison, the mean values are calculated and compared only for galaxies on the SFMS.

\subsection{Comparison of basic properties of galaxies in CO-VG and CO-CS}
\label{sec:comparison-samples}

\begin{figure}
\includegraphics[width=0.5\textwidth]{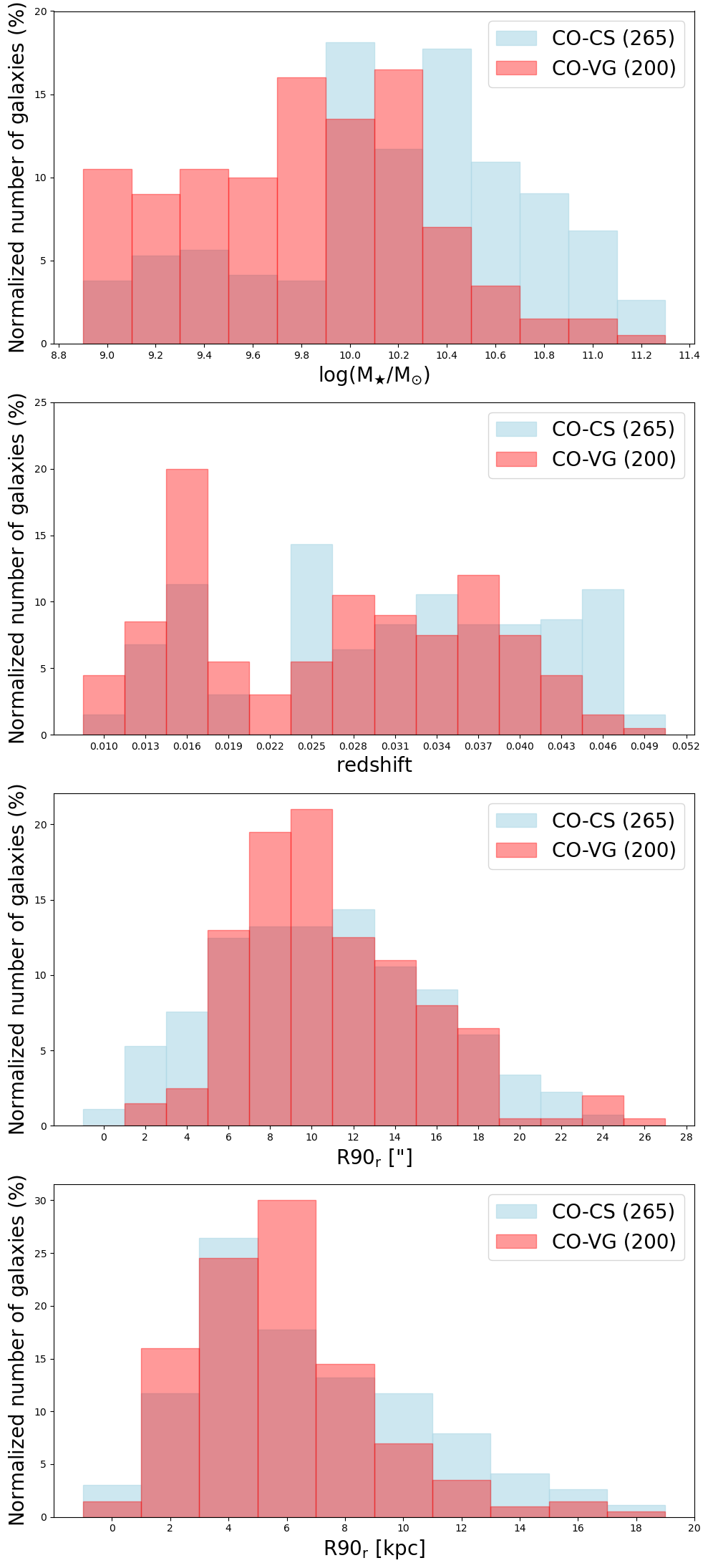} \\
\caption{Normalised histograms for the CO-VG (red) and the CO-CS (blue) to compare form top to the bottom: the redshift, stellar masses, and sizes (R$_{90,r}$, in arcsec and in kpc).}
\label{sample-histo}
\end{figure}

Figure \ref{sample-histo} presents a comparison between the CO-VG sample and the CO-CS, examining their redshift, stellar masses, and sizes (defined by R$_{90,r}$; which is the radius enclosing 90 percent of the Petrosian flux in the $r$ band from SDSS, see Table \ref{tab:properties}). There is a reasonable agreement in the redshift distributions of the two samples, with the CO-VG sample exhibiting a slight excess of nearby ($ z \lesssim 0.023$) objects and the CO-CS sample having more distant ($ z \gtrsim 0.043$) galaxies. The distributions in apparent size are also very similar, with a slight excess of small (R$_{90,r}  \lesssim 7''$) galaxies in the CO-CS sample and more medium-sized (R$_{90,r} \sim 10''$) galaxies in CO-VG. In terms of physical size as indicated in kiloparsecs (kpc), there is a slight excess of larger galaxies in the CO-CS sample than in the CO-VG sample. The largest difference is in the distribution of stellar mass. The CO-VG sample contains more than a factor of 2 more low-mass galaxies (log\,\Mstar/M$_{\odot}$ $<$ 10), than the CO-CS sample, while the CO-CS sample contains a larger fraction of high-mass (log\,\Mstar/M$_{\odot}$ $>$ 10.5) galaxies. Such a difference has previously been identified in studies of void samples and also results from hydrodynamical simulations (e.g. Illustris TNG, EAGLE), mainly  due to the fact that the star formation histories (SFHs) of void galaxies tend to be slower than those exhibited by galaxies in denser environments \citep{Alpaslan2015, Chen2017, Laigle2018, Rosas-Guevara2022, dominguez23-bbb, Rodriguez-Medrano2024}. It is therefore necessary to analyse the two samples taking into account the stellar mass as a crucial parameter. Furthermore, it is not possible to draw any statistically significant conclusions for the highest mass bins (log\,\Mstar $\gtrsim$ 11) due to the low number of galaxies in the CO-VG sample.

\subsection{SFMS}
\label{sec:SFMS}

As our reference, we employed the SFMS  established in \citet[][their Eq. 1]{janowiecki20}. Following them, we adopt $\pm$ 0.3 dex to define the width of the SFMS. Also, we employ their division line (-1.55 dex below the SFMS) to separate the TZ and the RS.

\begin{figure*}
   \includegraphics[width=\textwidth]{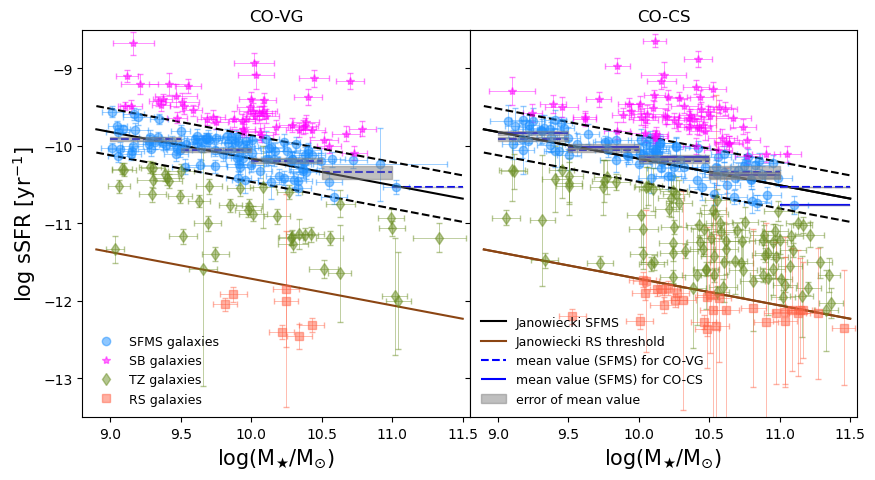}
     \caption{Specific star formation rate as a function of the stellar mass for the CO-VG sample (left) and the CO-CS sample (right). The SFMS is represented as a solid black line, and has been derived in \citet[][their Eq. 1]{janowiecki20}. The dashed black lines represent the $\pm$ 0.3 dex offset to the SFMS \citep{janowiecki20}. The brown solid line represent the threshold of the RS defined by \citep[][SFMS $<$ 1.55 dex]{janowiecki20}. Galaxies are colour coded according to their positions in the SFMS: i) pink stars represent starburst (SB) galaxies that lie above the SFMS, ii) blue dots star-forming (SF) galaxies within the SFMS ($\pm$ 0.3 dex), iii) green diamonds for galaxies that lie in the transition-zone (TZ), and iv) orange squares for red-sequence (RS) galaxies below the TZ (SFMS $<$ 1.55 dex). The mean sSFR per \Mstar\ bin is shown with a blue-dashed line for the CO-VG and blue-solid line for the CO-CS considering only those galaxy that belong to the SFMS. The extent of this line represent the width of the stellar mass bin. The error of the mean value is represented with a gray shadow.}    
\label{sSFR}
\end{figure*}

Figure \ref{sSFR} shows a decreasing correlation between the sSFR and \Mstar\ for both the CO-VG sample and the comparison sample. The mean values of the CO-CS sample are slightly higher than expected from the SFMS, which is probably due to the fact that we excluded void, cluster and AGN galaxies and therefore changed the CO-CS sample with respect to the sample used to derive the SFMS in \cite{janowiecki20}. In any case, the differences are very small (see Table \ref{tab:SFMS-comp}). 

For SF galaxies, the mean values of sSFR are, for most mass bins, slightly lower for  the CO-VG sample than for the CO-CS sample; the difference is, however, not significant (see Table~\ref{tab:sSFR}). For SF galaxies, the distribution of CO-VG galaxies is similar to the CO-CS sample. Above the SFMS, in the SB regime, the CO-VG sample is well distributed between log~\Mstar/\Msun = 9.0 – 10.5, while for the CO-CS sample the SB galaxies are more concentrated towards the mass bin log\Mstar/\Msun = 10.0 – 10.5.  Below the SFMS, the CO-VG galaxies are on average much closer to the SFMS than in the comparison sample, and the CO-CS sample is also more populated at higher stellar masses. Almost all CO-VG galaxies below the SFMS are in the TZ region, only a few galaxies are below this limit, in the RS region, which is mostly due to the selection of the objects (see Sect.~\ref{subsec:iram-sample}).

We note that the statistical reliability of the results of the two highest mass bins for this and all the subsequent relations studied in this work is very limited due to the low number of galaxies in each bin, in particular for SF galaxies. In the mass bin of log \Mstar/\Msun = 10.5--11.0, there are only 5 galaxies in the CO-VG sample and in the adjacent mass bin (log \Mstar/\Msun = 11.0--11.5) there is only one galaxy in each sample, so that the results in this bin are only illustrative. 

Since the distance to the SFMS is a parameter that characterizes the evolutionary path of galaxies, we calculate, following \citet[][their Eq. 3]{janowiecki20}, the distance to the SFMS, \distSFMS, as:

\begin{equation}
\label{eq:dist_SFMS}
\Delta{\rm SFMS} = \log\frac{\rm SFR}{ M_\star}  - \log\frac{\rm SFR_{\rm MS}}{ M_\star}, 
\end{equation}
where SFR$_{\rm MS}$ is the SFR corresponding to the SFMS value for a given \Mstar. In Table \ref{tab:SFMS-comp} we present the results of the mean values and standard deviation for the $\Delta{\rm SFMS}$, for SB, SFMS, TZ in both samples, the CO-VG and the CO-CS. 

\begin{table*}[!h]
\centering
\caption{\label{tab:sSFR} Star formation rate for CO-VG and CO-CS galaxies: Mean values for galaxies on the SFMS.}
\begin{tabular}{cccccccccc}
\hline
\noalign{\smallskip} \hline \noalign{\medskip}
\multicolumn{10}{c}{log sSFR [yr$^{-1}$]}  \\
\noalign{\smallskip} \hline \noalign{\medskip}
log \Mstar [\Msun] & \multicolumn{2}{c}{CO-VG} &  & \multicolumn{2}{c}{CO-CS} &  & $\Delta$ mean & $\sigma$  & KS \\
\cline{2-3} \cline{5-6}
range               & $n/n_{up}$& mean               &  & $n/n_{up}$ & mean    &  &      &      & \\
(1)                 &  (2)      &  (3)               &  & (4)        &  (5)    &  &  (6) & (7)  &  \\
\noalign{\smallskip} \hline \noalign{\medskip}
9.0  - 9.5  & 28/0 & -9.91  $\pm$ 0.03 &  & 21/0 & -9.83  $\pm$ 0.03 &  & -0.08 $\pm$ 0.04 & -1.90 &  0.060 \\
9.5  - 10.0 & 34/0 & -10.05 $\pm$ 0.03 &  & 11/0 & -10.01 $\pm$ 0.03 &  & -0.04 $\pm$ 0.04 & -0.94 &  0.640 \\
10.0 - 10.5 & 32/0 & -10.19 $\pm$ 0.03 &  & 29/0 & -10.15 $\pm$ 0.03 &  & -0.04 $\pm$ 0.04 & -0.95 &  0.453 \\
10.5 - 11.0 & 5/0 & -10.34  $\pm$ 0.09 &  & 23/0 & -10.38 $\pm$ 0.04 &  & 0.04  $\pm$ 0.09 & 0.39  &  0.636  \\
11.0 - 11.5 & 1/0 & -10.53  &  & 1/0  & -10.76  &  & 0.23  & -    &   - \\
\hline
\hline
\end{tabular}
\tablefoot{(1) Stellar mass range in the bin. (2) and (4) $n$: Number of galaxies for the CO-VG and the CO-CS samples, respectively, in the bin. $n_{\rm up}$: Number of upper limits in the mass bin. (3) and (5) Mean logarithm of the sSFR and its error of the CO-VG galaxies and the CO-CS galaxies, respectively, in the mass bin. (6) Difference of the mean logarithm of the sSFR between the CO-VG and the comparison sample and its error. (7) $\sigma$ = $\Delta$mean/err($\Delta$mean). (8) $p$-value of the Kolmogorov-Smirnov test.}
\end{table*}

\begin{table*}[!h]
\centering
\caption{\label{tab:SFMS-comp} <$\Delta$SFMS> measurements for CO-VG and CO-CS galaxies.}
\begin{tabular}{ccccccc}
\hline
\noalign{\smallskip} \hline \noalign{\medskip}
\multicolumn{7}{c}{<$\Delta$SFMS>}  \\
\noalign{\smallskip} \hline \noalign{\medskip}
     &      & CO-VG &      &       & CO-CS&        \\
\noalign{\smallskip} \hline \noalign{\medskip}
 & $n/n_{up}$ & <$\Delta$SFMS> & STD & $n/n_{up}$ & <$\Delta$SFMS> & STD \\
(1) &  (2)       &  (3)              & (4)      &  (5)       & (6)               & (7)\\
\noalign{\smallskip} \hline \noalign{\medskip}
SB   & 51/0 & 0.60  $\pm$ 0.03  & 0.24 & 61/0 & 0.64 $\pm$ 0.04 & 0.27 \\
SFMS & 100/0& 0.04  $\pm$ 0.02  & 0.16 & 85/0 & 0.06 $\pm$ 0.02 & 0.16 \\
TZ   & 42/0 & -0.75 $\pm$ 0.06  & 0.36 & 88/0 &-0.87 $\pm$ 0.04 & 0.37  \\
\hline
\hline
\end{tabular}
\tablefoot{(1) Position in relation to the SFMS; above the SFMS (SB), on the SFMS (SFMS) or below the SFMS, down to SFMS - 1.55 dex (TZ). (2) and (5) $n$: Total number of galaxies. $n_{up}$: Number of upper limits. (3) and (6) <$\Delta$SFMS> mean value and its error for the CO-VG and the CO-CS sample, respectively. (4) and (7) The standard deviation (STD) value for the CO-VG and the comparison sample, respectively.}
\end{table*}

\subsection{Molecular gas mass}

\begin{figure*}
\centering
   \includegraphics[width=\textwidth]{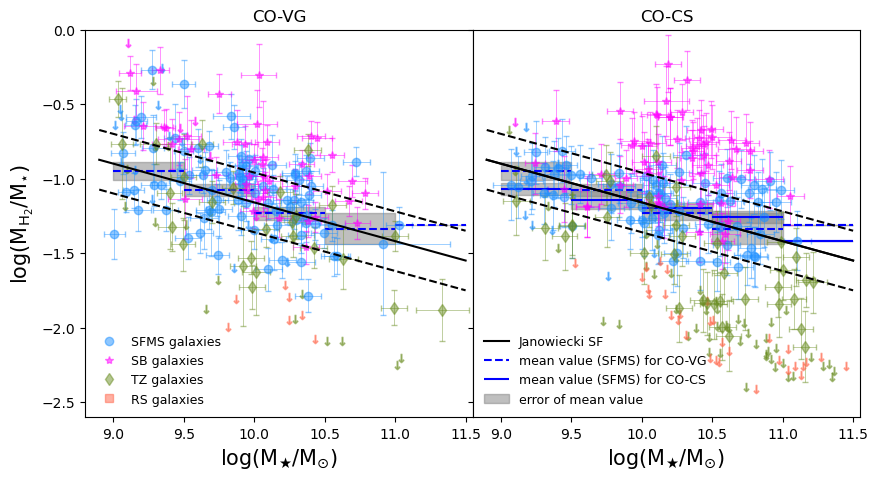}
     \caption{Molecular gas mass fraction as a function of the stellar mass for the CO-VG sample (left) and the CO-CS sample (right). The H$_2$MS from \cite{janowiecki20} is represented as a solid black line. The dashed black lines represent the $\pm$ 0.2 dex offset to the H$_2$MS. Galaxies are colour coded according to their positions in the SFMS as described in Fig.~\ref{sSFR}. The mean logarithm of the molecular gas mass fraction per \Mstar\ bin is shown with a blue-dashed line for the CO-VG and blue-solid line for the CO-CS considering only those galaxies that belong to the SFMS. The extent of this line represent the width of the stellar mass bin. The error of the mean value is represented with a gray shadow. The M$_{\rm H2}$ upper limits are represented with downward arrows.}
     \label{MH2_Mstar_Mstar}
\end{figure*}

Figure \ref{MH2_Mstar_Mstar} shows the relation between the molecular gas mass fraction, \mhtwo/\Mstar, and \Mstar\ for both the CO-VG sample and the comparison sample. Similarly to the SFMS, we adopted the H$_2$ main sequence, H$_2$MS, relation from \citet[][their Eq. 5]{janowiecki20} and identify galaxies falling within $\pm$ 0.2 dex of this relation as belonging to the  H$_2$MS. This relation was derived as a linear fit to the molecular gas mass fraction of the galaxies on the SFMS. \citet{janowiecki20} classified galaxies according to their gas content as gas-normal (within the H$_2$MS) as gas-rich (above this relation), or gas-poor (below this relation). Following their Eq. 7 we define the distance to the H$_2$MS as:

\begin{equation}
\label{eq:dist_H2MS}
\Delta{\rm H_2MS} = \log\frac{M_{H_2}}{ M_\star}  - \log\frac{M_{H_2, MS}}{ M_\star}, 
\end{equation}
where $M_{\rm H_2, MS}$ is the \mhtwo\ corresponding to the H$_2$MS value for a given \Mstar. 

There are no significant differences in the mean values of the molecular gas mass fraction in different mass bins for SFMS galaxies in the CO-VG and those in the CO-CS (see Table \ref{tab:MH2_Mstar}). We note that the lowest mass bin has a high number of upper limits, making the results less reliable.

\begin{table*}[h]
\centering
\caption{\label{tab:MH2_Mstar} Molecular gas mass fraction for CO-VG and CO-CS galaxies: Mean values for galaxies on the SFMS}
\begin{tabular}{cccccccccc}
\hline
\noalign{\smallskip} \hline \noalign{\medskip}
\multicolumn{10}{c}{log (\mhtwo/\Mstar)}  \\
\noalign{\smallskip} \hline \noalign{\medskip}
log(\Mstar) [\Msun] & \multicolumn{2}{c}{CO-VG} &  & \multicolumn{2}{c}{CO-CS} &  &   $\Delta$ mean & $\sigma$ &  KS  \\
\cline{2-3} \cline{5-6}
range               & $n/n_{up}$& mean               &  & $n/n_{up}$ & mean         &  &    &      &  \\
(1)                 &  (2)      &  (3)               &  & (4)        &  (5)         &  & (6)& (7)  & \\
\noalign{\smallskip} \hline \noalign{\medskip}
9.0  - 9.5  & 28/11 & -0.95 $\pm$ 0.06 &  & 21/8 & -1.07 $\pm$ 0.04 &  & 0.12  $\pm$ 0.07 & 1.68  & 0.010 \\
9.5  - 10.0 & 34/3  & -1.08 $\pm$ 0.04 &  & 11/1 & -1.15 $\pm$ 0.06 &  & 0.07  $\pm$ 0.07 & 0.91  & 0.263 \\
10.0 - 10.5 & 32/2  & -1.23 $\pm$ 0.04 &  & 29/1 & -1.20 $\pm$ 0.04 &  & -0.03 $\pm$ 0.06 & -0.60 & 0.932 \\
10.5 - 11.0 & 5/0   & -1.34 $\pm$ 0.11 &  & 23/0 & -1.26 $\pm$ 0.04 &  & -0.08 $\pm$ 0.11 & -0.69 & 0.525 \\
11.0 - 11.5 & 1/0   & -1.31  &  & 1/0  & -1.42  &  & 0.10  & -    &  - \\
\hline
\hline
\end{tabular}
\tablefoot{(1) Stellar mass range in the bin. (2) and (4) $n$: Number of galaxies for the CO-VG and CO-CS sample, respectively in the bin. $n_{up}$: Number of upper limits in the bin. (3) and (5) Mean logarithm of the molecular gas mass fraction and its error of the CO-VG galaxies and the CO-CS galaxies, respectively in the bin. (6) Difference of the mean logarithm of the molecular gas mass fraction between the CO-VG sample and the comparison and its error. (7) $\sigma$ = $\Delta$mean/err($\Delta$mean). (8) $p$-value of the Kolmogorov-Smirnov test.}
\end{table*}

\begin{figure*}
\centering
   \includegraphics[width=\textwidth]{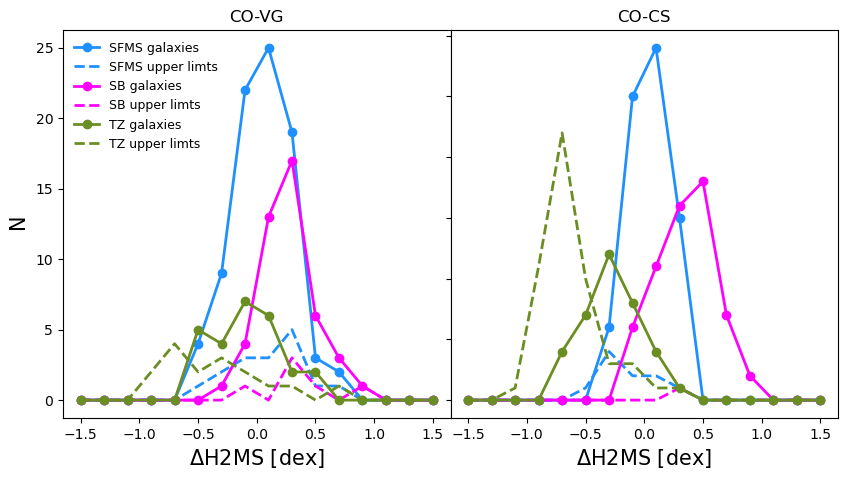}
     \caption{Histograms of $\Delta$H$_2$MS for the CO-VG sample (left) and the comparison sample (right). The three categories of galaxies are colour coded with respect to their positions with  the SFMS; i) pink for SB galaxies above the SFMS, ii) blue for SF galaxies within the SFMS ($\pm$ 0.3 dex), and iii) green for galaxies that lie in the TZ. H$_{2}$ detections are shown as full lines and  non-detections as dotted histograms.}
     \label{H2MS-histo}
\end{figure*}

In Fig. \ref{H2MS-histo}, we present a histogram of the distributions of $\Delta$H$_2$MS for three categories (SB, SFMS, and TZ) in both the CO-VG sample and the comparison sample. The mean and the median values, as well as the standard deviation for the $\Delta$H$_2$MS distributions, are summarised in Table \ref{tab:H2-histo}. The three distributions exhibit an overlap, especially in the case of the CO-VG sample. 
The distribution of the SFMS galaxies are similar for CO-CS and CO-VG galaxies, with no significant difference neither in the mean nor in the standard deviation of $\Delta$H$_2$MS. For the SB galaxies, there is, however, a tentative difference with the mean value of $\Delta$H$_2$MS for CO-CS being shifted to higher values (by 0.07 dex, corresponding to 2$\sigma$) compared to CO-VG. For the TZ galaxies, we find a similar trend, in the sense that the mean value for CO-VG galaxies is higher (i.e. closer to the SFMS value) than for the CO-CS sample by 0.27 dex (corresponding to 4$\sigma$). The trend for this galaxy group has, however, to be taken with caution because of the sparse sampling of the CO-VG sample, in particular the lack of massive galaxies in the TZ, and the large number of upper limits. 
 
\begin{table*}[!h]
\centering
\caption{\label{tab:H2-histo} <$\Delta$H$_2$MS> measurements for CO-VG and CO-CS galaxies.}
\begin{tabular}{ccccccc}
\hline
\noalign{\smallskip} \hline \noalign{\medskip}
\multicolumn{7}{c}{<$\Delta$H$_2$MS>}  \\
\noalign{\smallskip} \hline \noalign{\medskip}
     &      & CO-VG &      &       & CO-CS&        \\
\noalign{\smallskip} \hline \noalign{\medskip}
 & $n/n_{up}$ & <$\Delta$H$_2$MS> & STD & $n/n_{up}$ & <$\Delta$H$_2$MS> & STD \\
(1) &  (2)       &  (3)              & (4)      &  (5)       & (6)               & (7)\\
\noalign{\smallskip} \hline \noalign{\medskip}
SB   & 51/6    & 0.28  $\pm$ 0.03   & 0.23  &  61/1   & 0.35 $\pm$ 0.03   & 0.26 \\
SFMS & 100/16  & 0.06  $\pm$ 0.03   & 0.26  &  85/10  & 0.01 $\pm$ 0.02   & 0.20  \\
TZ   & 32/16   & -0.20 $\pm$ 0.06   & 0.37  &  88/52  & -0.47 $\pm$ 0.03  & 0.30 \\
\hline
\hline
\end{tabular}
\tablefoot{(1) Position according the SFMS; above (SB), on (SFMS) or in the CO-voids-transition zone (TZ). (2) and (5) $n$: Total number of galaxies. $n_{up}$: Number of upper limits. (3) and (6) <$\Delta$H$_2$MS> mean value and its error for the CO-VG and the CO-CS sample, respectively. (4) and (7) The standard deviation (STD) value for the CO-VG and the comparison sample, respectively.}
\end{table*}

\begin{figure*}
\centering
   \includegraphics[width=\textwidth]{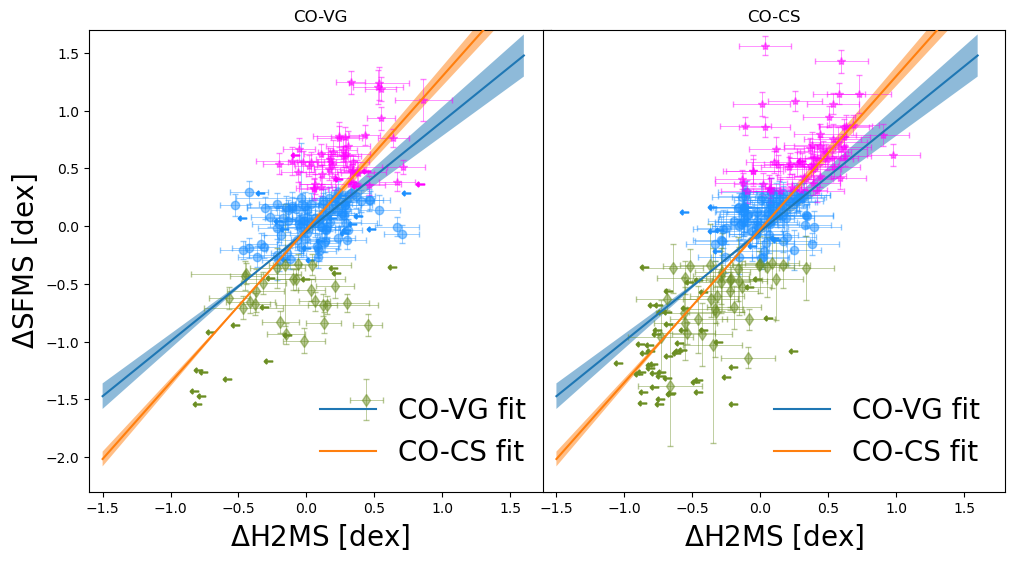}
     \caption{$\Delta$SFMS as a function of the $\Delta$H$_2$MS for the CO-VG sample (left) and the comparison sample (right). Galaxies are colour coded according to their positions in the SFMS as described in Fig.~\ref{sSFR}. The blue and orange lines represent fits for the CO-VG sample and the comparison sample, respectively. The respective blue and orange shadows represent the 1 $\sigma$ dispersion of the fits. The CO non-detections are shown with leftward arrows.}
     \label{SFMSD-H2MSD}
\end{figure*}

Apart from its dependence on \Mstar, the molecular gas properties also depend on the distance to the SFMS, $\Delta$SFMS \citep[see][]{saintonge16, tacconi18, janowiecki20}. Figure \ref{SFMSD-H2MSD} shows the relation between $\Delta$SFMS and $\Delta$H$_2$MS. There is a clear relation between both quantities visible for the comparison sample. For the CO-VG sample, this relation has a higher dispersion and a smaller dynamic range in $\Delta$H$_2$MS. We used the linear least squares method to fit the data in both samples as $\Delta$SFMS  = $a + m\,\times\, \Delta$H$_2$MS. We derived a slope of $m = 0.95 \pm 0.09$ and $1.33 \pm 0.06$  and $a = -0.03 \pm 0.26 $ and $-0.05 \pm 0.03$ for the CO-VG sample and CO-CS, respectively. The correlation coefficient is $r = 0.59$ and $0.83$ for the CO-VG sample and CO-CS, respectively. 

In conclusion, we find no major differences in the dependence of the molecular gas fraction on stellar mass for galaxies on the SFMS between galaxies in the CO-VG and in the CO-CS. However, when considering the distribution of $\Delta$H$_2$MS and its relation to $\Delta$SFMS, we find a positive correlation for both the CO-VG and the CO-CS sample, in the sense that the larger the distance in SFR from the SFMS, the larger the distance in the molecular gas mass fraction from the H$_2$MS. However, this relation is tighter for galaxies in the CO-CS than for those in the CO-VG sample, for which the dynamical range in the molecular gas mass fraction seems also to be smaller. 

\subsection{Star formation efficiency (SFE = SFR/$M_{\rm H_2}$)}

\begin{figure*}
\centering
   \includegraphics[width=\textwidth]{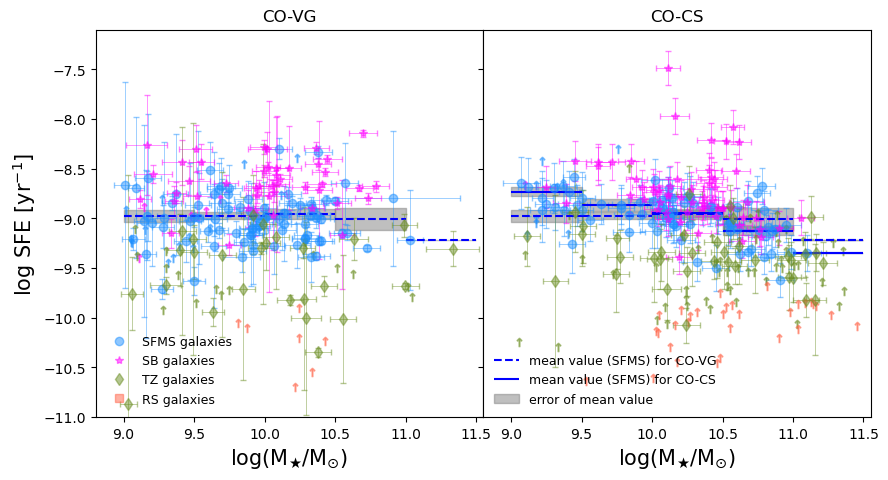}
     \caption{Star formation efficiency (SFE) as a function of the stellar mass for CO-VG sample (left) and CO-CS (right) sample. Galaxies are colour coded according to their position in relation to the SFMS as described in Fig.~\ref{sSFR}. The mean logarithm of the SFE per \Mstar\ bin is shown with a blue-dashed line for the CO-VG and blue-solid line for the CO-CS considering only those galaxies that belong to the SFMS. The extent of this line represent the width of the stellar mass bin. The error of the mean value is represented with a gray shadow. The lower limits are shown with upward arrows.}
     \label{SFE_Mstar}
\end{figure*}

Figure \ref{SFE_Mstar} shows the relation between the SFE and the stellar mass. For the SFMS galaxies in the comparison sample, a slightly decreasing trend is visible. This trend is not seen for the CO-VG sample, for which the relation seems to be constant across all mass bins. The different trends are also seen when comparing the mean values for the individual bins (Table~\ref{tab:SFE}). The low value of the KS test (p-value = 0.002) for the lowest mass bin confirms the difference in the trends at low stellar masses for the two samples. These trends remain the same when calculating the mean of all galaxies in each mass bin (i.e. taking together SB, SFMS and TZ galaxies), thus they seem to be robust and a property of the different environments, independent of SF activity.

\begin{table*}[h]
\centering
\caption{\label{tab:SFE} Star formation efficiency for CO-VG and CO-CS galaxies: Mean values for galaxies on the SFMS.}
\begin{tabular}{cccccccccc}
\hline
\noalign{\smallskip} \hline \noalign{\medskip}
\multicolumn{10}{c}{log SFE [yr$^{-1}$]}  \\
\noalign{\smallskip} \hline \noalign{\medskip}
log(\Mstar) [\Msun] & \multicolumn{2}{c}{CO-VG} &  & \multicolumn{2}{c}{CO-CS} &  &   $\Delta$ mean & $\sigma$  & KS \\
\cline{2-3} \cline{5-6}
range               & $n/n_{low}$& mean               &  & $n/n_{low}$ & mean         &  &    &      &  \\
(1)                 &  (2)      &  (3)               &  & (4)        &  (5)          &  & (6)&(7) & \\
\noalign{\smallskip} \hline \noalign{\medskip}
9.0  - 9.5  & 28/11 & -8.97 $\pm$ 0.06 &  & 21/8 & -8.73 $\pm$ 0.05 &  & -0.24 $\pm$ 0.08 & -3.18 & 0.002 \\
9.5  - 10.0 & 34/3  & -8.97 $\pm$ 0.04 &  & 11/1 & -8.87 $\pm$ 0.06 &  & -0.10 $\pm$ 0.07 & -1.44 & 0.085 \\
10.0 - 10.5 & 32/2  & -8.96 $\pm$ 0.05 &  & 29/1 & -8.95 $\pm$ 0.04 &  & -0.01  $\pm$ 0.06 & -0.11 & 0.453 \\
10.5 - 11.0 & 5/0   & -9.01 $\pm$ 0.11 &  & 23/0 & -9.12 $\pm$ 0.05 &  & 0.11  $\pm$ 0.12 & 0.97  & 0.710 \\
11.0 - 11.5 & 1/0   & -9.22  &  & 1/0  & -9.35  &  & 0.13   & -    &  - \\
\hline
\hline
\end{tabular}
\tablefoot{(1) Stellar mass range in the bin. (2) and (4) $n$: Number of galaxies for the CO-VG and CO-CS galaxies respectively in the bin. $n_{low}$: Number of lower limits in the bin. (3) and (5) Mean logarithm of the star formation efficiency and its error of the CO-VG and the comparison sample, respectively in the bin. (6) Difference of the mean logarithm of the star formation efficiency between the CO-VG sample and the comparison sample and its error. (7) $\Delta$mean/err($\Delta$mean). (8) $p$-value of the Kolmogorov-Smirnov test.}
\end{table*}

\begin{figure*}
\centering
   \includegraphics[width=\textwidth]{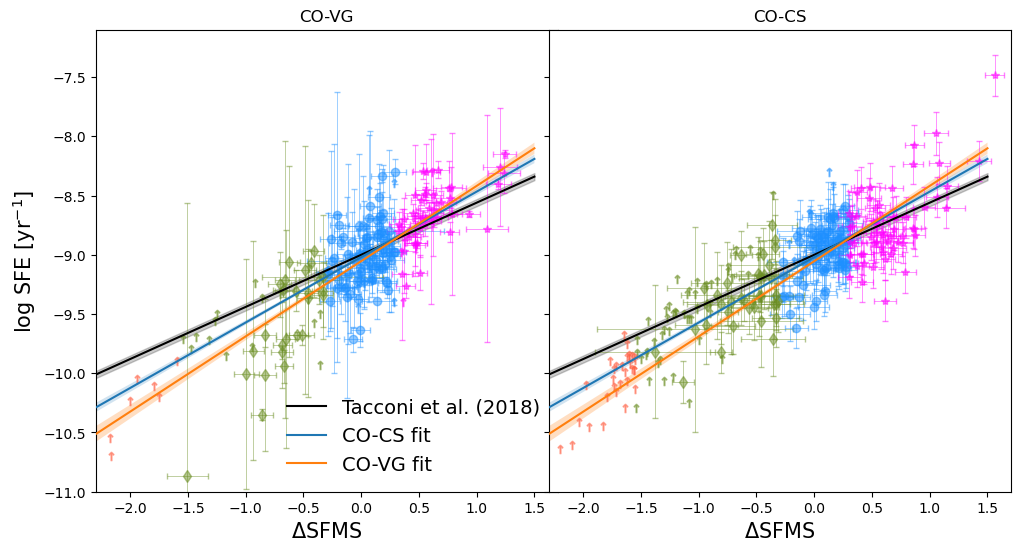}
     \caption{Star formation efficiency as a function of the SFMS distance for the CO-VG sample (left) and the comparison sample (right). Galaxies are colour coded according to their position in relation to the SFMS as described in Fig.~\ref{sSFR}. The blue and orange lines represent fits for the CO-VG sample and the comparison sample, respectively, considering the four categories of galaxies. The respective blue and orange shadows represent the 1 $\sigma$ dispersion of the fits. The black thick line represent the power law found by \cite{tacconi18}.} 
     \label{SFE_SFMSD_2}
\end{figure*}

Figure \ref{SFE_SFMSD_2} shows the relation between the SFE and  $\Delta$SFMS for both the CO-VG and the CO-CS samples. A linear relationship is visible for both samples. 
We carried out a linear least square fit to the data in both samples, as log SFE = $a + m\,\times\, \Delta$SFMS.
We derived a slope of $m = 0.64 \pm 0.03$ and $0.55 \pm 0.02 $  for the CO-VG sample and CO-CS, respectively, and $a = -9.0 \pm 0.1$ for both samples. For comparison, Fig.~\ref{SFE_SFMSD_2} also incorporates the relationship reported by \cite{tacconi18} for a large sample of galaxies at different redshifts (log SFE $\propto$ $\Delta$SFMS$^{0.44}$. Here, we arbitrarily adjust the abscissa to best fit our data). The slope derived for the CO-VG sample is steeper than for the comparison sample, and also steeper than the slope of \cite{tacconi18}, suggesting a tendency of a high SFE of local void galaxies above the SFMS.

\subsection{Line intensity ratio (R$_{21}$ = I$_{CO(2-1)}$/I$_{CO(1-0)}$)}
\label{sec:line-ratio}

\begin{figure}
\centering   \includegraphics[width=0.5\textwidth]{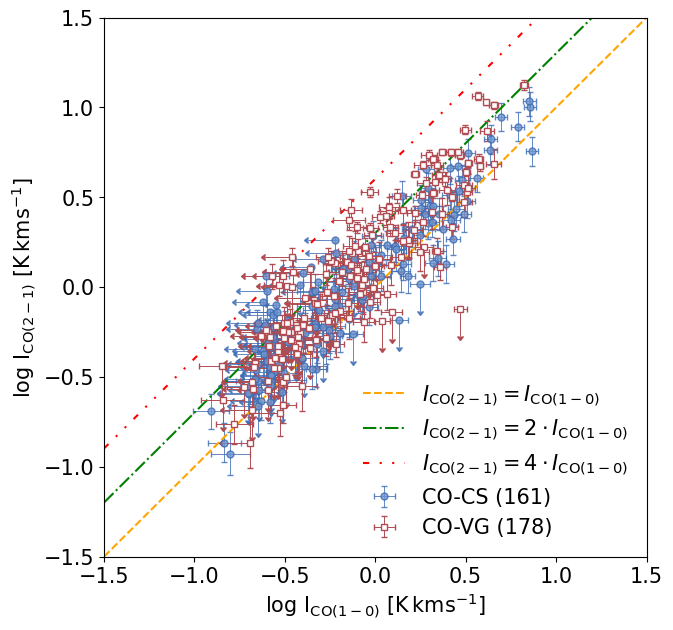}
     \caption{\icotwo\ and \icoone\ line emission  (on the T$_{mb}$ temperature scale) for the CO-VG sample and the CO-CS sample. To guide the eye we include lines for fixed ratios between \icotwo\ and \icoone.}
     \label{R21}
\end{figure}

Figure \ref{R21} shows the correlation between the observed velocity integrated intensities \icoone\ and \icotwo\ for both samples. The observations of the CO(1-0) and CO(2-1) emission were carried out with different beam-sizes. Therefore, in order to interpret the ratio between them, \rtwoone = \icotwo/\icoone, one has to consider, apart from the excitation temperature of the gas, two main parameters: the source size relative to the beam and the opacity of the molecular gas (see e.g. \citealt{solomon97}, Sect. 3.1.1, or \citealt{dominguez22}, Appendix A). Following \citet{dominguez22}, we can derive from their Eq. A7 the expected values of \rtwoone\ for optically thick gas in thermal equilibrium in the case of a point-like or a very extended source: For emission with a point-like distribution we expect a ratio \rtwoone = $\ensuremath{(\bar{T}_{Bc2-1}/\bar{T}_{Bc1-0}})(\Theta^2_{\rm B(CO1-0)}/\Theta^2_{\rm B(CO2-1)}$ (with $\bar{T}_{Bc}$ being the central intrinsic brightness ratio and $\Theta_{B}$ the FWHM of the respective beams). On the other hand, for a source that is more extended than the beams, or for beam-matched observations with $\ensuremath{\Theta_{\rm B(CO1-0)} = \Theta_{\rm B(CO2-1)}}$, we expect \rtwoone $\ensuremath{  = \bar{T}_{Bc2-1}/\bar{T}_{Bc1-0}}$. 

In order to interpret \rtwoone\ for our observations with different beam sizes, we need to take the ratio between source and beam size into account. In Fig. \ref{R21-R90} we therefore show the relation between \rtwoone\ and $R_{90,r}$, which parametrizes the source size. We consider two cases separately in order to be able to calculate the mean values with the Kaplan-Meier estimator which does not allow to mix upper and lower limits: i) Subsamples including galaxies with detection and upper limits in \rtwoone, i.e. data points derived from detections \icoone\ and detections or upper limits in \icotwo. ii) Subsamples including galaxies with detection and lower limits in \rtwoone, i.e. data points derived from detections \icotwo\ and detections or upper limits in \icoone. We include in the figures, lines showing the theoretically expected \rtwoone\ ratios for different intrinsic central brightness temperature ratios, following \citet{dominguez22}. These theoretical ratios assume that the distribution of the CO emission decreases exponentially with a scale length of $0.2\times r_{\rm 25}$ (see Sect.~{\ref{sec:molecular_gas_mass}). In addition, the mean values for different $R_{90,r}$ bins for both samples are plotted. 

We find for both samples that \rtwoone\ is decreasing with $R_{\rm 90,r}$, as theoretically expected. The values of \rtwoone\ for both samples are compatible within the errors for most of the bins that contain a significant ($n \gtrsim$ 10) number of galaxies. Only in the radial bin 15\arcsec $< R_{\rm 90,r} < $ 20\arcsec, there is a higher \rtwoone\ (by $\sim 3\sigma$) for CO-CS galaxies. In both samples, the mean values of \rtwoone\ follows reasonably well the predicted relation for an intrinsic central brightness temperature ratio of $\bar{T}_{Bc2-1}/\bar{T}_{Bc1-0} \sim 0.7$. This value is consistent with optically thick gas with a temperature of 5-10~K \citep[$\bar{T}_{Bc2-1}/\bar{T}_{Bc1-0}$ = 0.6 corresponds to gas with a temperature of $\sim$ 5~K, 0.8 to $\sim$ 10~K, and 0.9 to $\sim$ 21~K, and higher excitation temperatures yield a line ratio $\sim$ 1,][]{braine92,leroy09}.

The value of $\bar{T}_{Bc2-1}/\bar{T}_{Bc1-0} \sim 0.7$ agrees well with results for beam-matched line ratios, \rtwoone\, from other galaxy samples which indicate that the ratio in disks is typically \rtwoone $\sim 0.5-0.8$, with higher values ($\sim 1$) in the centres \citep{casoli91, leroy09}. In agreement with this, \citet{braine93} found an average line ratio of 0.89 $\pm$ 0.6 for the central regions of a small sample of nearby spiral galaxies. Furthermore, \citet{braine92} found a small difference in their sample for isolated galaxies (\rtwoone = 0.82) and perturbed objects (\rtwoone = 1.11).  \citet{casasola15} found \rtwoone $\sim 0.6 - 0.9$ for four low-luminosity AGNs from the NUGA survey. Also for the Milky Way typical line ratios of $\sim 0.6 - 0.8$ in the disk and $\sim$1 toward the Galactic centre were found \citep{Sakamoto1995, Oka1996, Oka1998}.

\begin{figure*}
\centerline{
\includegraphics[width=0.5\textwidth]{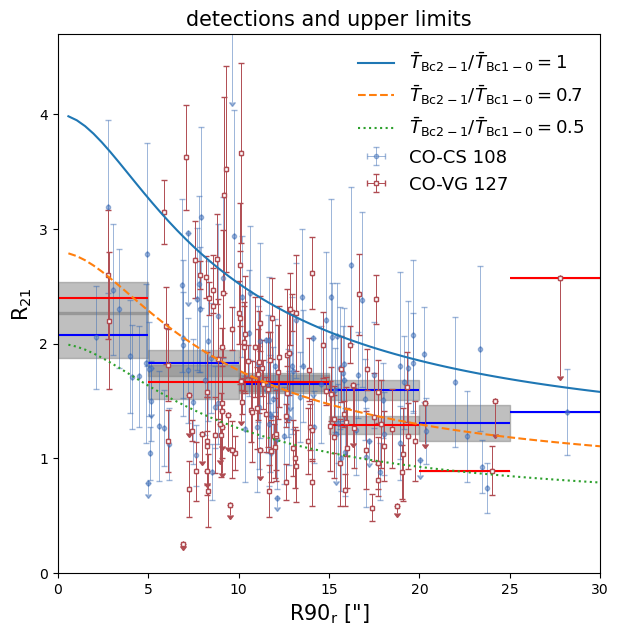}
\includegraphics[width=0.5\textwidth]{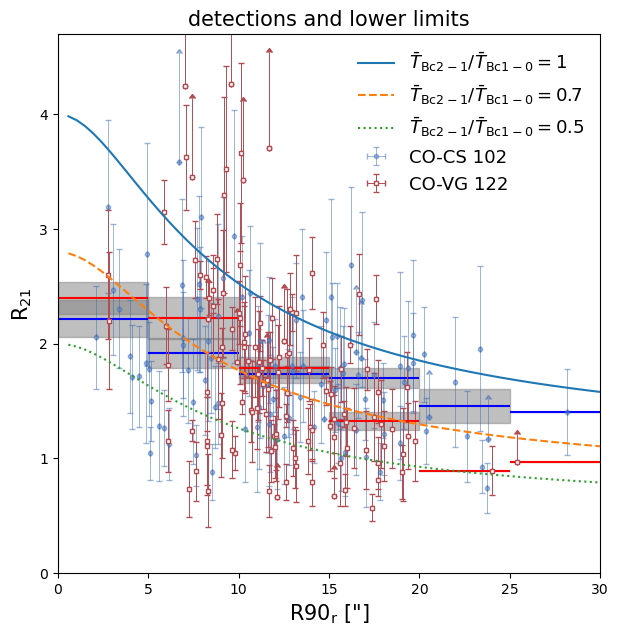}
}
\caption{Observed line intensity ratio R$_{21}$ = I$_{CO(2-1)}$/I$_{CO(1-0)}$ as a function of the r-Petrosian R$_{90,r}$. Left panel: Only detections and upper limits (i.e. data points with non-detections in I$_{CO(1-0)}$ were not considered). Right panel: Only detections and lower limits (i.e. data points with non-detections in I$_{CO(2-1)}$ were not considered.}
\label{R21-R90}
\end{figure*}

\begin{table*}[h]
\centering
\caption{\label{tab:R21_1} CO(2--1)-to-CO(1--0) line ratio, \rtwoone, for CO-VG and CO-CS galaxies: Mean values}
\begin{adjustbox}{max width=\textwidth}
\begin{tabular}{cccccccccc}
\hline
\noalign{\smallskip} \hline \noalign{\medskip}
\multicolumn{10}{c}{R$_{\rm 21} = I_{\rm CO(2-1)} / I_{\rm CO(1-0)}$, including detections and upper limits (i.e. only galaxies with  $I_{\rm CO(1-0)} > 3\sigma$)}  \\
\noalign{\smallskip} \hline 
\noalign{\smallskip} \hline \noalign{\medskip}
R$_{90,r}$ & \multicolumn{2}{c}{CO-VG} &  & \multicolumn{2}{c}{CO-CS} &  &   $\Delta$ mean & $\sigma$  & KS  \\
\cline{2-3} \cline{5-6}
range               & $n/n_{up}$& mean               &  & $n/n_{up}$ & mean         &  &    &   &      \\
(1)                 & (2)       & (3)                &  & (4)        & (5)          &  & (6)&(7) & (8) \\ 
\noalign{\smallskip} \hline \noalign{\medskip}
0.0  - 5.0  & 2/0  & 2.40 $\pm$ 0.14 &  & 10/1 & 2.07 $\pm$ 0.20 &  & 0.33  $\pm$ 0.24 & 1.34  & 0.455   \\
5.0  - 10.0 & 39/9 & 1.66 $\pm$ 0.15 &  & 28/3 & 1.83 $\pm$ 0.12 &  & -0.17 $\pm$ 0.19 & -0.88 & 0.143  \\
10.0 - 15.0 & 51/2 & 1.66 $\pm$ 0.08 &  & 32/5 & 1.65 $\pm$ 0.08 &  & 0.02  $\pm$ 0.11 & 0.16  & 0.476   \\
15.0 - 20.0 & 31/2 & 1.29 $\pm$ 0.08 &  & 29/3 & 1.60 $\pm$ 0.09 &  & -0.30 $\pm$ 0.12 & -2.55 & 0.015   \\
20.0 - 25.0 & 3/2  & 0.89 $\pm$ 0.00 &  & 8/1  & 1.31 $\pm$ 0.16 &  & -0.42 $\pm$ 0.16 & -2.61 & 0.836  \\
25.0 - 30.0 & 1/1  & 2.57  &  & 1/0  & 1.41  &  & 1.17  & -    & - \\
\noalign{\smallskip} \hline \noalign{\medskip}
\multicolumn{9}{c}{R$_{\rm 21} = I_{\rm CO(2-1)} / I_{\rm CO(1-0)}$, including detections and lower limits (i.e. only galaxies with  $I_{\rm CO(2-1)} > 3\sigma$)}  \\
\noalign{\smallskip} \hline \noalign{\medskip}
R$_{90,r}$ & \multicolumn{2}{c}{CO-VG} &  & \multicolumn{2}{c}{CO-CS} &  &   $\Delta$ mean & $\sigma$  & KS  \\
\cline{2-3} \cline{5-6}
 range              & $n/n_{low}$& mean     &  & $n/n_{low}$ & mean  &  & &   \\
(1)                 & (2)       & (3)                &  & (4)        & (5)          &  & (6)&(7) & (8) \\ 
\noalign{\smallskip} \hline \noalign{\medskip}
0.0  - 5.0  & 2/0  & 2.40 $\pm$ 0.14 &  & 9/0  & 2.22 $\pm$ 0.16 &  & 0.18  $\pm$ 0.22 & 0.86 & 0.545 \\
5.0  - 10.0 & 34/5 & 2.22 $\pm$ 0.18 &  & 26/1 & 1.92 $\pm$ 0.13 &  & 0.31  $\pm$ 0.22 & 1.37 & 0.472 \\
10.0 - 15.0 & 54/5 & 1.79 $\pm$ 0.10 &  & 29/2 & 1.73 $\pm$ 0.07 &  & 0.06  $\pm$ 0.12 & 0.49 & 0.433 \\
15.0 - 20.0 & 30/1 & 1.33 $\pm$ 0.08 &  & 28/2 & 1.70 $\pm$ 0.09 &  & -0.38 $\pm$ 0.12 & -3.11 & 0.004 \\
20.0 - 25.0 & 1/0  & 0.89 $\pm$ 0.00 &  & 9/2  & 1.46 $\pm$ 0.15 &  & -0.57 $\pm$ 0.15 & -3.75 & 0.400\\
25.0 - 30.0 & 1/1  & 0.97  &  & 1/0  & 1.41  &  & -0.44 & -    & - \\
\hline
\hline
\end{tabular}
\end{adjustbox}
\tablefoot{(1) Range in Petrosian R$_{90,r}$. (2) and (4) $n$: Number of galaxies for CO-VG galaxies and CO-CS, respectively, in the R$_{90,r}$ range. $n_{up}$ ($n_{low}$): Number of upper (lower) limits of \rtwoone\ in the Petrosian R$_{90,r}$ range. (3) and (5) Mean of the R$_{21}$ ratio and its error of the CO-VG and the CO-CS sample, respectively, in the  R$_{90,r}$ range. (6) Difference of the R$_{21}$ ratio between the CO-VG and the CO-CS samples and its error. (7) $\sigma$ = $\Delta$mean/err($\Delta$mean). (8) $p$-value of the Kolmogorov-Smirnov test.}
\end{table*}

\section{Discussion}
\label{sec:discussion} 

Simulations and observations suggest that galaxy properties are strongly influenced by their large-scale environment. However, there is no clear consensus yet about the concrete effect that this has on the gas and the SF. While simulations generally predict a suppression of SF in present-day galaxies with decreasing distance from filaments \citep[e.g.][]{Hasan2023, Bulichi2024}, observations give opposing results. Some studies find an enhancement of the sSFR in filament galaxies \citep{Vulcani2019, Darvish2014}, possibly caused by filaments assisting gas cooling and thereby enhancing star formation. However, within filaments, other phenomena with opposite consequences may also occur, such as gas-stripping which can quench star formation \citep[e.g.][]{Chen2017, Kraljic2018, Winkel2021}. In any case, these large-scale environmental processes are expected to mostly affect the atomic hydrogen (HI) supply of the galaxies \citep[see e.g.][]{Hasan2023}, and not directly the molecular gas which is formed from the HI in galaxies. Observations of HI will be important to search for differences between galaxies in voids and denser environments, and are in progress for the CAVITY project.

In the present study, we focus on observations of the molecular gas mass, to find out whether the amount of molecular gas or the SFE depends on the large-scale environment, and, in particular, if there are differences between galaxies in voids and in filaments and walls. Our large sample ($200$ void galaxies in CO-VG and a similar number of galaxies in filaments and walls in CO-CS) allows a solid statistical comparison between both environments. In order to provide a reliable comparison, we made sure that the relevant parameters of this study (SFR, molecular gas mass and stellar mass) were derived consistently in both samples and, in addition, in order to prevent any ambiguities, we excluded galaxies with AGNs from our study. The two most important internal parameters that control the molecular gas mass fraction and the SFE are the stellar mass and the distance to the SFMS. Accordingly, we based our comparison on these two parameters.

With respect to the molecular gas mass fraction, M$_{\rm H_2}$/\Mstar, we did not find significant differences between CO-VG and CO-CS across all mass bins for the SFMS galaxies. This finding is consistent with the results of \cite{dominguez22}, who analysed similar properties in a pilot sample of $\sim 20$ void galaxies. When considering also SB and TZ galaxies, we found a tentative indication that the dynamic range of M$_{\rm H_2}$/\Mstar\ is higher in the comparison sample than in the CO-VG sample, suggesting that the environment could affect the molecular gas content in galaxies. 

On the other hand, a much clearer difference is observed in the variation of the SFE with respect to stellar mass between galaxies in CO-VG and in CO-CS (see Fig.~\ref{SFE_Mstar}). For the comparison sample, the mean SFE exhibits a decline of approximately 5$\%$ from the highest to the lowest value when considering only the most populated first four mass bins (log \Mstar/\Msun, = 9.0 -- 11.0). This trend aligns with prior investigations \citep[see, e.g.][for the highest mass galaxies]{saintonge17, lisenfeld23}. Quite differently, the SFE-stellar mass relationship in the CO-VG sample remains practically constant in this mass range. The clearest difference is in the lowest mass bin (log \Mstar/\Msun, = 9.0 -- 9.5) where the SFE in CO-VG lies 0.24 dex (corresponding to $>3\sigma$, and KS $p$-value $\sim$ 0.002) below the value of the comparison sample. This means that at low stellar masses, SF is less efficient in void galaxies compared to filaments and walls. 

What could be the cause of such a difference? 
A variation of the SFE might be associated to a change in the properties of the molecular gas. We note that molecular gas measured by CO(1-0) probes relatively low-density gas ($\sim3000$ cm$^{-3}$), part of which might not be associated with Giant Molecular Clouds (GMCs) able to form stars. Other molecules, such as  HCN(1-0) with higher critical densities are better tracer for the star-forming molecular gas \citep[e.g.][]{gao04, Gracia-Carpio2006, Garcia-Burillo2012}. Thus, the variation of the fraction of high-density molecular gas to low-density molecular gas, that is unable to form stars, across the SFMS might be the reason for the observed variation of the SFE in CO-CS \citep[see][]{saintonge17}, and if it is the case, such a variation might be absent in the void galaxies. The reason for such a variation --or lack of it-- is, however, not obvious from the single-dish data. Resolved data are needed to shed more light on this question. Resolved studies relying on CO and HCN \citep{Usero2015, Bigiel2016, Jiminez-Donaire2019} have shown that indeed the dense molecular gas fraction depends on many galactic properties as the stellar or gas surface density, molecular-to-atomic mass ratio and the dynamical equilibrium pressure. Surprisingly, the ratio between SFR and dense (i.e. HCN traced) molecular gas varies considerably - more than the CO-traced SFE-- with galactic environmental conditions. Studies at the scale of individual GMC in different galaxies have shown that their properties are driven by their galactic environment on kpc scale \citep{Sun2022}, and that the pressure of the hydrostatic ISM plays a role in shaping their density and velocity dispersion \citep{Hughes2013}. Thus, it might be possible that these effects play a role for the (small) differences that we have found between CO-VG and CO-CS galaxies, most likely due to effects that the large-scale intergalactic environment might have on the galactic ISM. High-resolution observations with ALMA are currently under way for the CAVITY sample in order to make progress in this question.

Another internal mechanism responsible for quenching, known as "morphological quenching," has been proposed (\cite{Martig2009}; \cite{Bluck2014}). This suggests that the presence of a bulge can stabilise the gas and inhibit star formation. Simulations have confirmed that this relation has a causal basis through dynamical effects \citep[``dynamical quenching'',][]{gensior20} of the central spheroids that increase the gas velocity dispersion and stabilizes the region against SF. This mechanism may help to explain the observed decrease in the SFE with stellar mass, given that the bulge fraction also increases with stellar mass. If this mechanism were to explain the difference found in the SFE between the void and the comparison sample - in particular the higher SFE at low masses for void galaxies -  a smaller bulge-to-disk ratio for low-mass galaxies in voids would be a necessary condition, and will need to be explored by further analysis of the optical CAVITY data.

\section{Conclusions}
\label{sec:conclusions}

In this paper, following the CO-CAVITY pilot project containing CO data for 20 void galaxies \citep{dominguez22}, we present data for the molecular gas mass of a large sample of void galaxies (200 galaxies, CO-VG sample) which we analyze and compare to a similar-sized comparison sample (265 galaxies, CO-CS sample). This analysis represents so far the largest statistical data set for the molecular gas in void galaxies and is aimed at studying the molecular gas content and properties in void galaxies and searching for possible differences compared to galaxies that inhabit denser structures. 

We present new data of the CO(1--0) and CO(2--1) line emission observed with the IRAM 30\,m IRAM telescope for a sample of 106 galaxies belonging to the CAVITY project \citep{Perez2024}. For CO(1--0), we recorded 64 detections, 9 tentative detections, and 33 non-detections, while for CO(2--1), we obtained 63 detections, 8 tentative detections, and 35 non-detections. 
In addition to the 106 CAVITY galaxies, we included a sample of 16 CO-VGS galaxies that were previously observed using the IRAM 30\,m telescope \citep{dominguez22}, and 89 xCOLD GASS void galaxies. Finally, we exclude 11 galaxies classified as AGN, thus the final CO-VG sample is composed of 200 galaxies. 

We selected the xCOLD GASS survey as our comparison sample, after removing void and cluster galaxies, as well as galaxies with AGNs. We conducted an analysis of the sSFR, molecular gas mass fraction, and the star formation efficiency as a function of stellar mass and distance to the SFMS. Below, we summarize the key conclusions.

\begin{enumerate}

\item The distribution of the molecular gas mass fraction as a function of \Mstar\ of galaxies on the SFMS  is overall consistent with the distribution of the comparison sample. For starburst galaxies, defined as objects located above the SFMS, the mean log(\mhtwo/\Mstar) is slightly lower in CO-VG than for CO-CS galaxies and for galaxies in the TZ defined as objects below the SFMS, it is higher, indicating that the distribution of log(\mhtwo/\Mstar) as a function of distance from the SFMS might be more concentrated than for the comparison sample.

\item  The mean values for the SFE for void galaxies on the SFMS show no trend with \Mstar\, in contrast to the comparison sample; for void galaxies, the mean value of SFE ranges between $\sim$ -8.94 and -8.99 whereas in the comparison sample SFE decreases from -8.73 to -9.12 from the lowest to the highest mass bin. These different trends might be related to differences in the properties of the molecular gas at the spatial scales of GMCs, possibly caused by the different environments. However, the exact cause is unclear and higher resolution observations are necessary to make progress.

\item The ratio of the CO(2-1)/CO(1-0) intensity ratio, \rtwoone, shows a decreasing trend with galactic radius, R$_{90,r}$, for both samples, being compatible, within the scatter, with an intrinsic brightness temperature ratio (i.e., beam-matched) ratio of \rtwoone $\sim$ 0.7. This ratio is consistent with optically thick, thermalized emission with an excitation temperature of 5-10 K, in agreement with other studies based on beam-matched observations.

\end{enumerate}

This is one of the first studies of molecular gas in a large number of void galaxies. Our analysis shows that there are no differences between void galaxies and galaxies within filaments and walls in terms of the total amount of molecular gas, but there are differences in the SFE, suggesting that the physical mechanisms affecting the SFE are different in void galaxies. Further data (IFU, high-resolution CO, HI), which forms part of the CAVITY project, will elucidate, in more detail, the distinctions between void galaxies and those within filaments and walls.

\begin{acknowledgements}
We thank the referee for very constructive comments improving the content and presentation of this paper. Thanks to the staff at IRAM Pico Veleta for their support during the observations. This work is based on observations carried out with the IRAM 30\,m telescope. IRAM is supported by INSU/CNRS (France), MPG (Germany) and IGN (Spain). We acknowledge financial support by the research projects AYA2017-84897-P, PID2020-113689GB-I00, and PID2020-114414GB-I00, financed by MCIN/AEI/10.13039/501100011033, the project A-FQM-510-UGR20 financed from FEDER/Junta de Andaluc\'{\i}a-Consejer\'{\i}a de Transformaci\'on Econ\'omica, Industria, Conocimiento y Universidades/Proyecto and by the grants P20-00334 and FQM108, financed by the Junta de Andaluc\'{\i}a (Spain). MIR acknowledge financial support from Grant AST22.4.4, funded by Consejer\'{\i}a de Universidad, Investigaci\'on e Innovaci\'on and Gobierno de Espa\~na and Uni\'on Europea – NextGenerationEU. Funding for this work/research was provided by the European Union (MSCA EDUCADO, GA 101119830). BB acknowledges financial support from the Grant AST22-4.4 funded by Consejer\'{\i}a de Universidad, Investigaci\'on e Innovaci\'on and Gobierno de Espa\~na and Uni\'on Europea – NextGenerationEU, and by the research project PID2020-113689GB-I00 financed by MCIN/AEI/10.13039/501100011033. DE acknowledges support from a Beatriz Galindo senior fellowship (BG20/00224) from the Spanish Ministry of Science and Innovation. J. F-B acknowledges support from the PID 2022-140869 NB-100 grant from the Spanish Ministry of Science and Innovation. GTR acknowledges financial support from the research project PRE2021-098736, funded by MCIN/AEI/10.13039/501100011033 and FSE+. KK gratefully acknowledges funding from the Deutsche Forschungsgemeinschaft (DFG, German Research Foundation) in the form of an Emmy Noether Research Group (grant number KR4598/2-1, PI Kreckel) and the European Research Council’s starting grant ERC StG-101077573 (“ISM-METALS"). LSM acknowledges support from Juan de la Cierva fellowship (IJC2019-041527-I). M.A-F. acknowledges support from the Emergia program (EMERGIA20-38888) from Consejer\'ia de Universidad, Investigaci\'on e Innovaci\'on de la Junta de Andaluc\'ia. M.S.P. and A.B. acknowledge the support of the Spanish Ministry of Science, Innovation and Universities through the project PID-2021-122544NB-C43. PVG acknowledges that the project that gave rise to these results received the support of a fellowship from “la Caixa” Foundation (ID 100010434). The fellowship code is B005800. SBD acknowledges financial support by the grant AST22.4.4, funded by Consejer\'{\i}a de Universidad, Investigaci\'on e Innovaci\'on and Gobierno de Espa\~na and Uni\'on Europea – NextGeneration EU, also funded by PID2020-113689GB-I00, financed by MCIN/AEI. RGB acknowledges financial support from the Severo Ochoa grant CEX2021-001131-S funded by MCIN/AEI/ 10.13039/501100011033 and grant PID2022-141755NB-I00. SDP acknowledges financial support from Juan de la Cierva Formaci\'on fellowship (FJC2021-047523-I) financed by MCIN/AEI/10.13039/501100011033 and by the European Union `NextGenerationEU'/PRTR, Ministerio de Econom\'ia y Competitividad under grants PID2019-107408GB-C44, PID2022-136598NB-C32, and is grateful to the Natural Sciences and Engineering Research Council of Canada, the Fonds de Recherche du Qu\'ebec, and the Canada Foundation for Innovation for funding. TRL acknowledges support from Juan de la Cierva fellowship (IJC2020-043742-I). This work made use of the following software packages: \texttt{astropy} \citep{astropy:2013, astropy:2018, astropy:2022}, \texttt{Jupyter} \citep{2007CSE.....9c..21P, kluyver2016jupyter}, \texttt{matplotlib} \citep{Hunter:2007}, \texttt{numpy} \citep{numpy}, \texttt{pandas} \citep{mckinney-proc-scipy-2010, pandas_10697587}, \texttt{python} \citep{python}, \texttt{scipy} \citep{2020SciPy-NMeth, scipy_10909890}, \texttt{astroquery} \citep{2019AJ....157...98G, astroquery_10799414} and \texttt{scikit-learn} \citep{scikit-learn, sklearn_api, scikit-learn_10666857}. Software citation information aggregated using \texttt{\href{https://www.tomwagg.com/software-citation-station/}{The Software Citation Station}} \citep{software-citation-station-paper, software-citation-station-zenodo}. This research has made use of the NASA/IPAC Extragalactic Database, which is funded by the National Aeronautics and Space Administration and operated by the California Institute of Technology. Funding for SDSS-III has been provided by the Alfred P. Sloan Foundation, the Participating Institutions, the National Science Foundation, and the U.S. Department of Energy Office of Science. The SDSS-III Web site is \url{http://www.sdss3.org/}. The SDSS-IV site is \url{http://www.sdss.org}.
\end{acknowledgements}


\bibliographystyle{aa}
\bibliography{main}

\appendix

\section{CO emission line spectra of the detected IRAM 30\,m CO-CAVITY galaxies.}
\label{app:co10_spectra}

Figures~\ref{co10-emission} and \ref{co21-emission} show the observed spectra of the CO(1--0) and CO(2--1) emission lines, respectively. 

\begin{figure*}
\centerline{
\includegraphics[width=4.5cm]{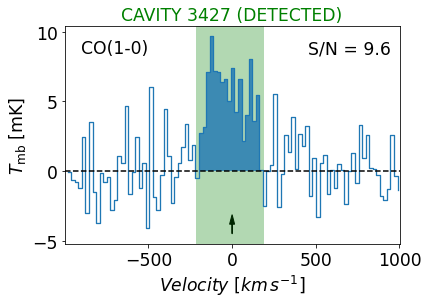}
\includegraphics[width=4.5cm]{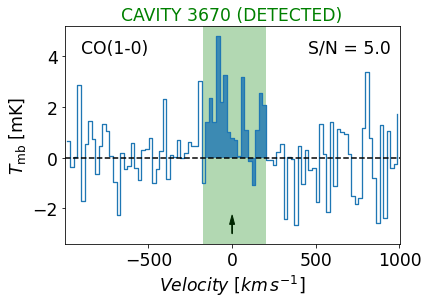}
\includegraphics[width=4.5cm]{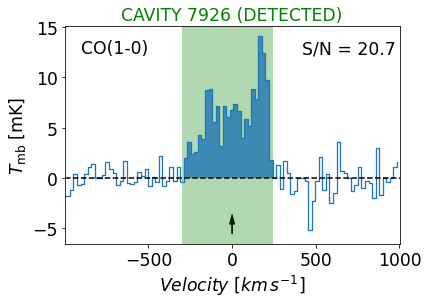}
\includegraphics[width=4.5cm]{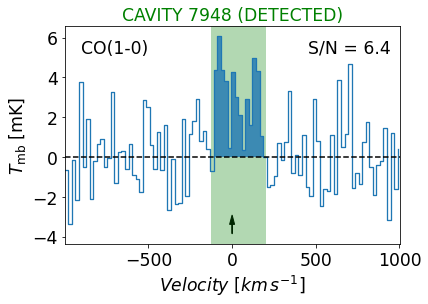}
}
\quad
\centerline{
\includegraphics[width=4.5cm]{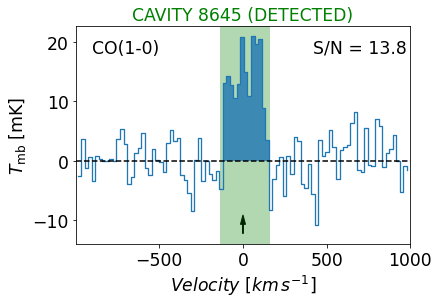}
\includegraphics[width=4.5cm]{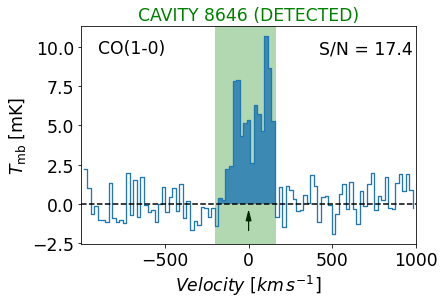}
\includegraphics[width=4.5cm]{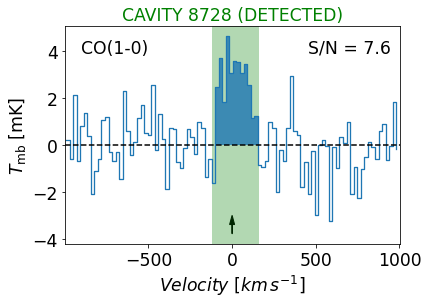}
\includegraphics[width=4.5cm]{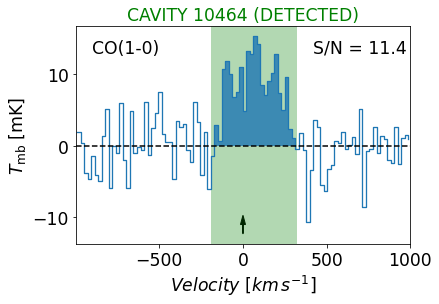}
}
\quad
\centerline{
\includegraphics[width=4.5cm]{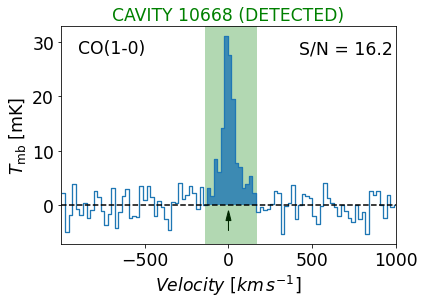}
\includegraphics[width=4.5cm]{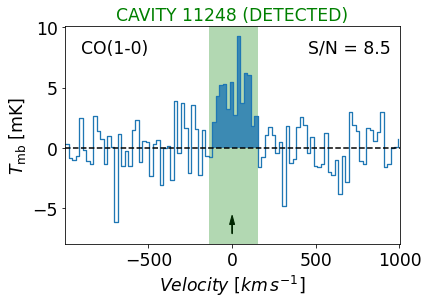}
\includegraphics[width=4.5cm]{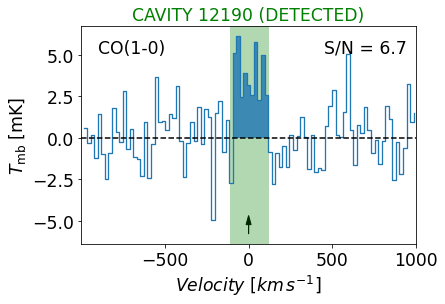}
\includegraphics[width=4.5cm]{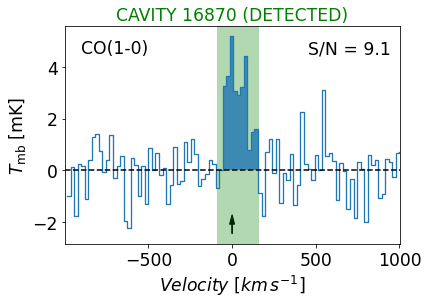}
}
\quad
\centerline{
\includegraphics[width=4.5cm]{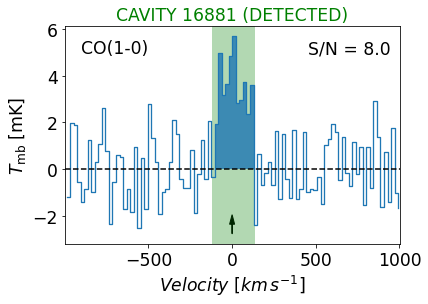}
\includegraphics[width=4.5cm]{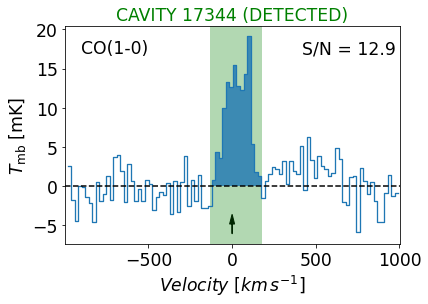}
\includegraphics[width=4.5cm]{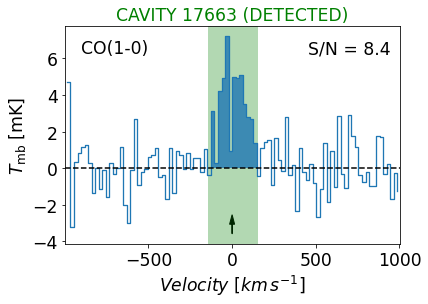}
\includegraphics[width=4.5cm]{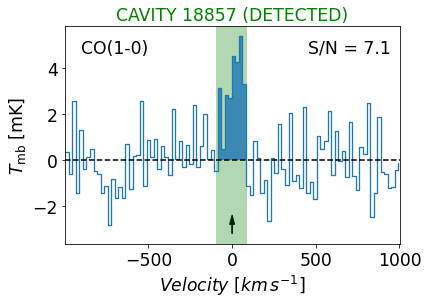}
}
\quad
\centerline{
\includegraphics[width=4.5cm]{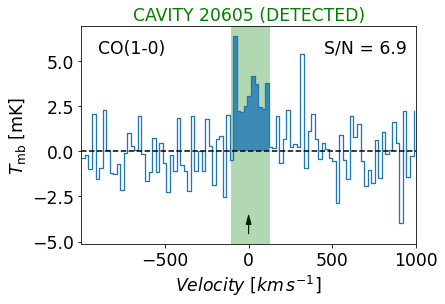}
\includegraphics[width=4.5cm]{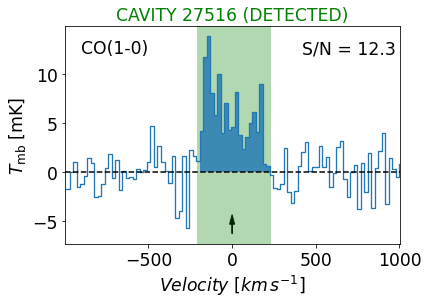}
\includegraphics[width=4.5cm]{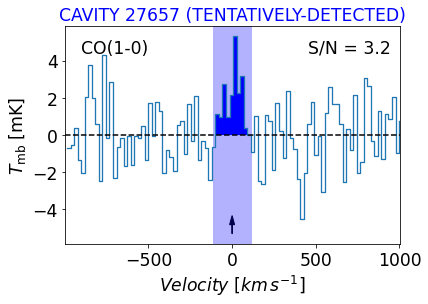}
\includegraphics[width=4.5cm]{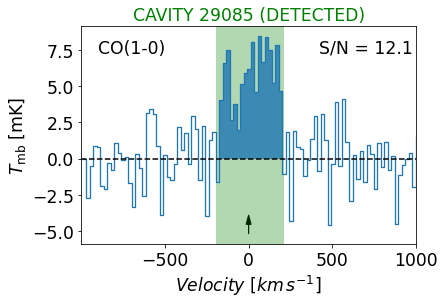}
}
\quad
\centerline{
\includegraphics[width=4.5cm]{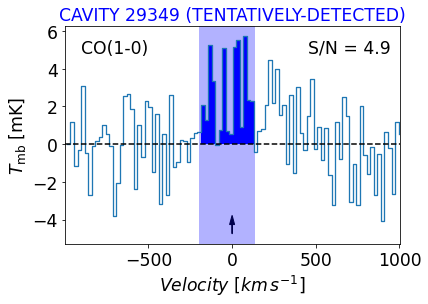}
\includegraphics[width=4.5cm]{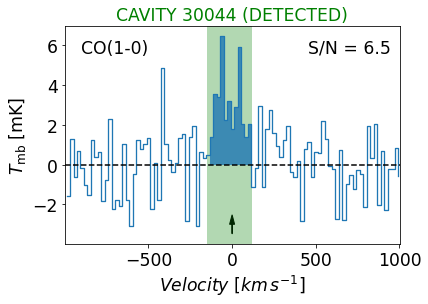}
\includegraphics[width=4.5cm]{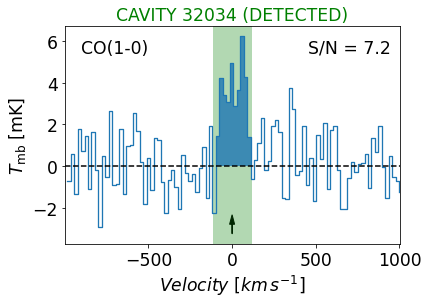}
\includegraphics[width=4.5cm]{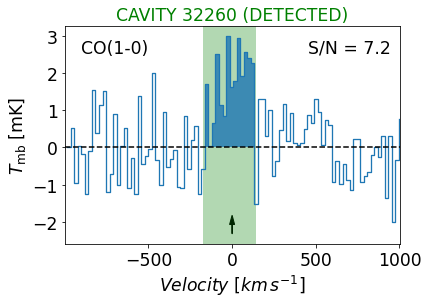}
}
\quad
\centerline{
\includegraphics[width=4.5cm]{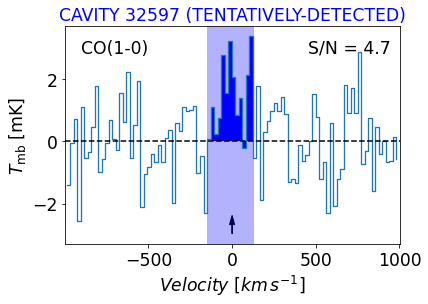}
\includegraphics[width=4.5cm]{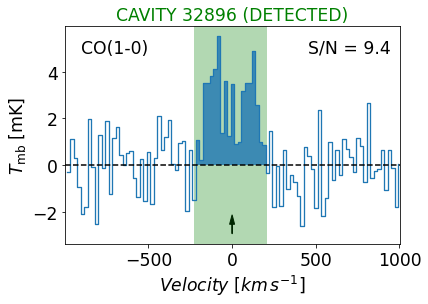}
\includegraphics[width=4.5cm]{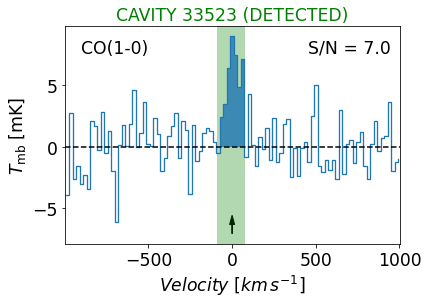}
\includegraphics[width=4.5cm]{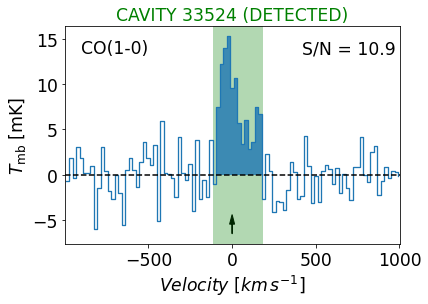}
}
\caption{CO(1--0) spectra of the void galaxies. The velocity resolution is 20 km s$^{-1}$. The zero velocity corresponds to the optical redshift $z_{\rm SDSS}$. The coloured shaded area represents the region over which the line is integrated to determine the total flux.}
\label{co10-emission}
\end{figure*}

\begin{figure*}
\label{fig:spectra-co10}
\centerline{
\includegraphics[width=4.5cm]{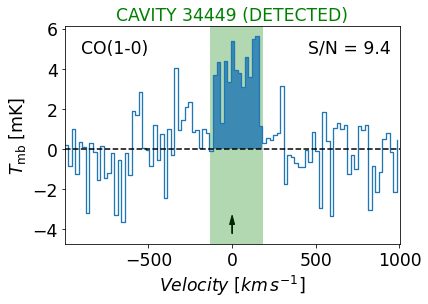}
\includegraphics[width=4.5cm]{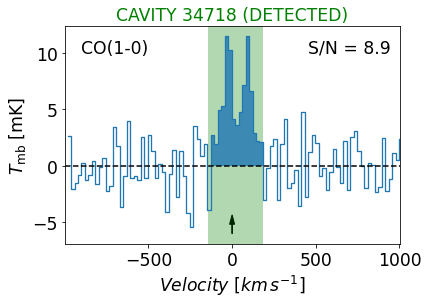}
\includegraphics[width=4.5cm]{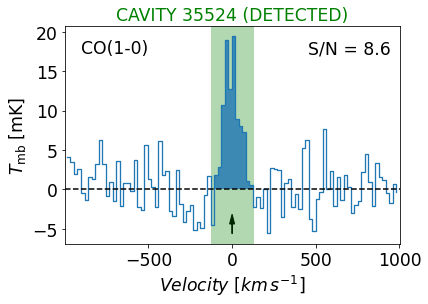}
\includegraphics[width=4.5cm]{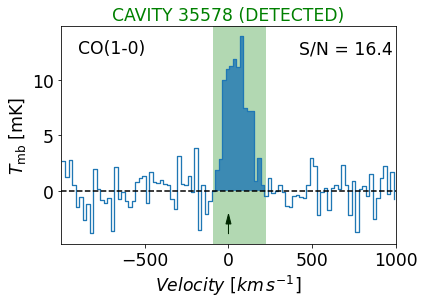}
}
\quad
\centerline{
\includegraphics[width=4.5cm]{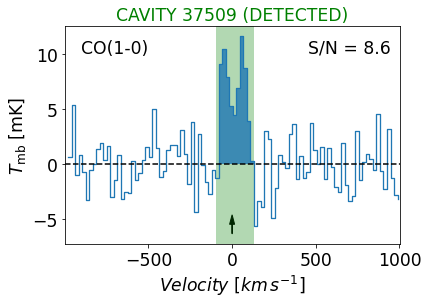}
\includegraphics[width=4.5cm]{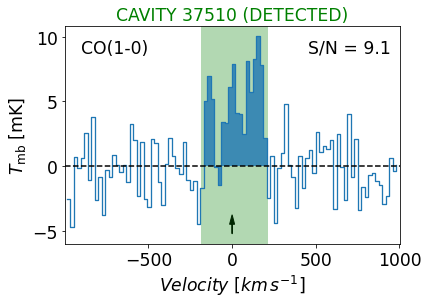}
\includegraphics[width=4.5cm]{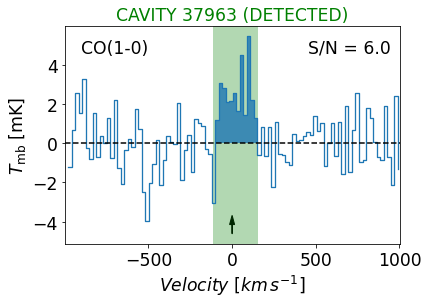}
\includegraphics[width=4.5cm]{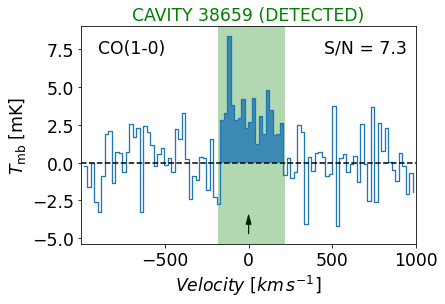}
}
\quad
\centerline{
\includegraphics[width=4.5cm]{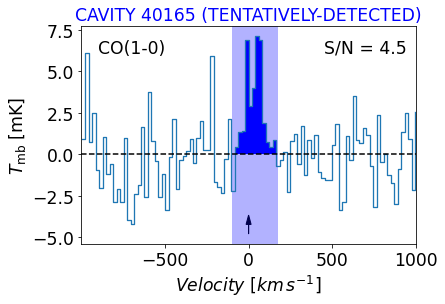}
\includegraphics[width=4.5cm]{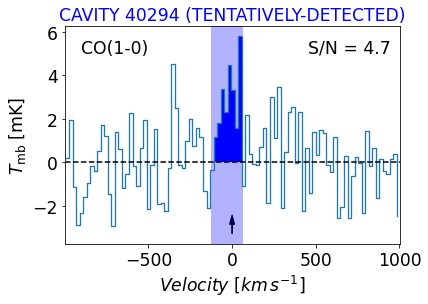}
\includegraphics[width=4.5cm]{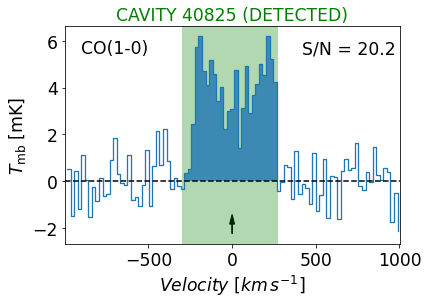}
\includegraphics[width=4.5cm]{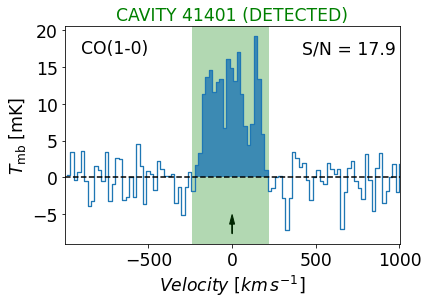}
}
\quad
\centerline{
\includegraphics[width=4.5cm]{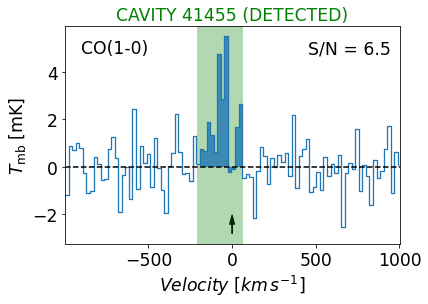}
\includegraphics[width=4.5cm]{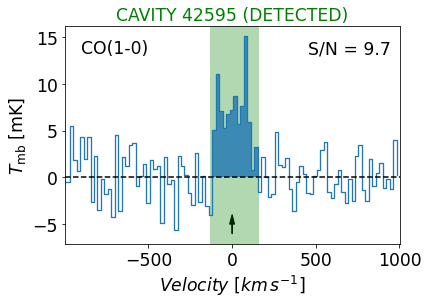}
\includegraphics[width=4.5cm]{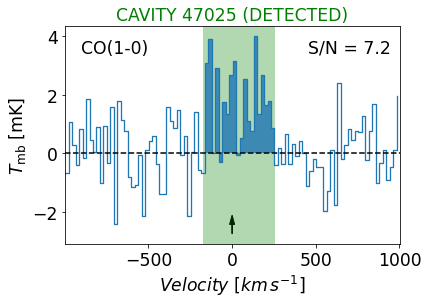}
\includegraphics[width=4.5cm]{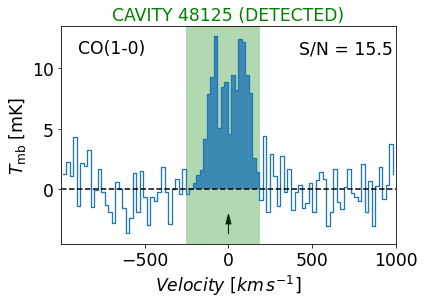}
}
\quad
\centerline{
\includegraphics[width=4.5cm]{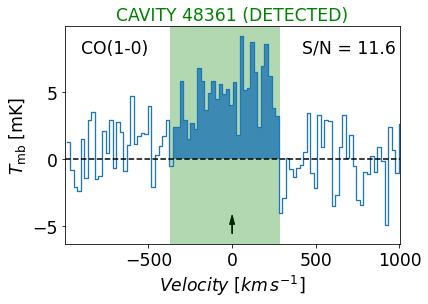}
\includegraphics[width=4.5cm]{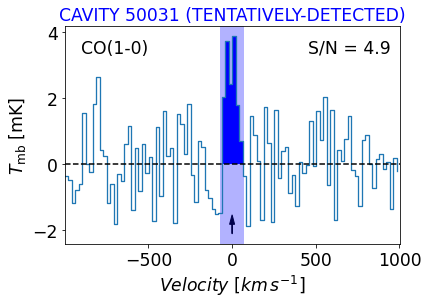}
\includegraphics[width=4.5cm]{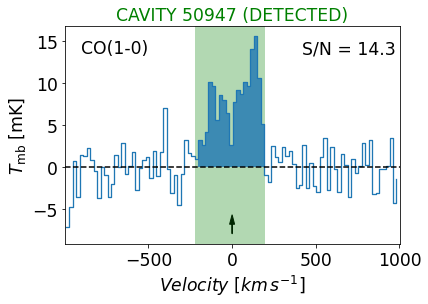}
\includegraphics[width=4.5cm]{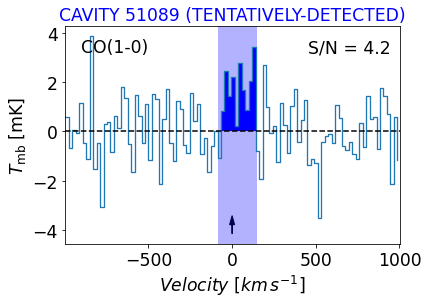}
}
\quad
\centerline{
\includegraphics[width=4.5cm]{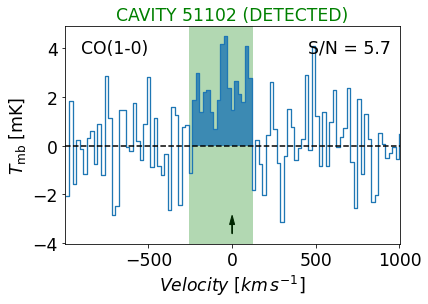}
\includegraphics[width=4.5cm]{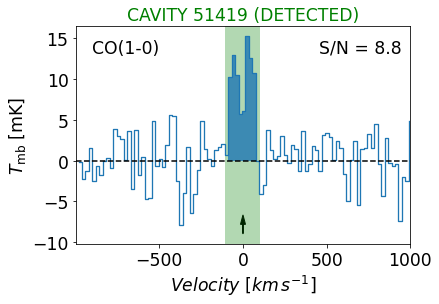}
\includegraphics[width=4.5cm]{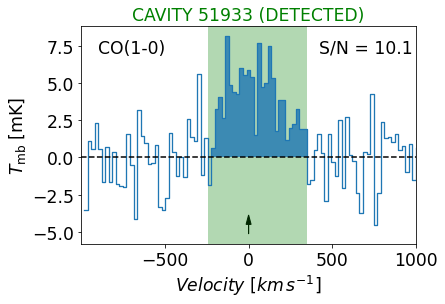}
\includegraphics[width=4.5cm]{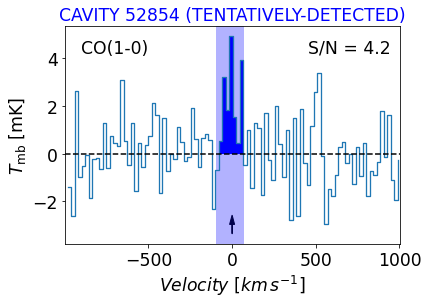}
}
\quad
\centerline{
\includegraphics[width=4.5cm]{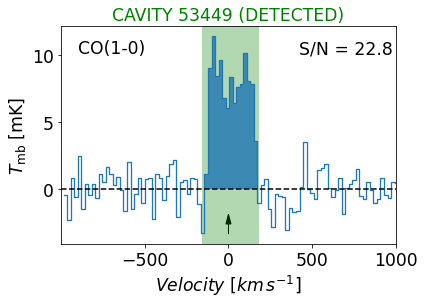}
\includegraphics[width=4.5cm]{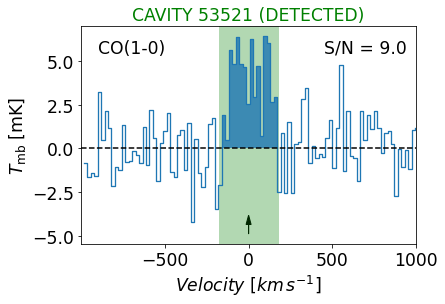}
\includegraphics[width=4.5cm]{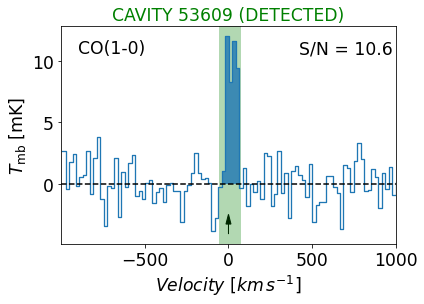}
\includegraphics[width=4.5cm]{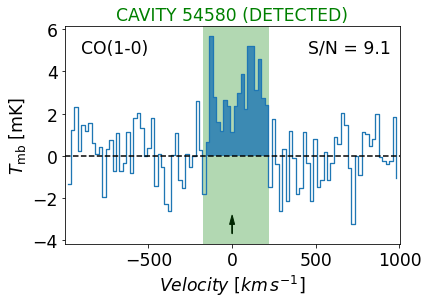}
}
\addtocounter{figure}{-1}
\caption{(continued)}
\end{figure*}

\begin{figure*}
\centerline{
\includegraphics[width=4.5cm]{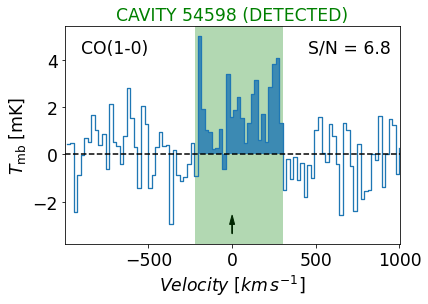}
\includegraphics[width=4.5cm]{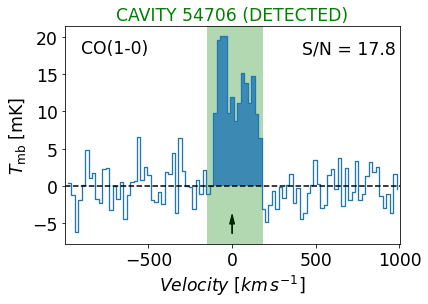}
\includegraphics[width=4.5cm]{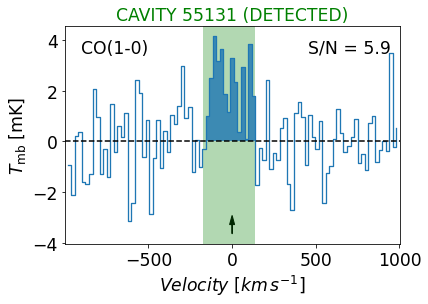}
\includegraphics[width=4.5cm]{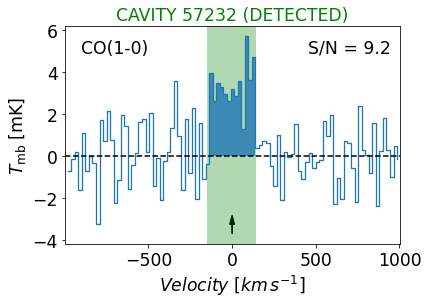}
}
\quad
\centerline{
\includegraphics[width=4.5cm]{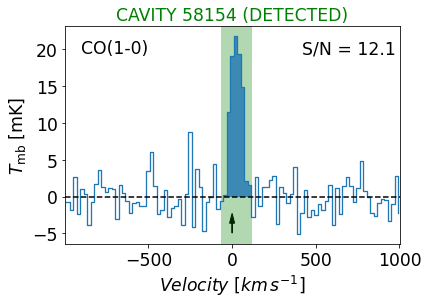}
\includegraphics[width=4.5cm]{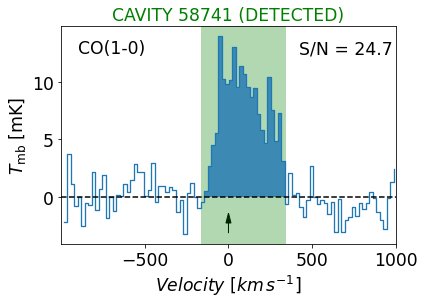}
\includegraphics[width=4.5cm]{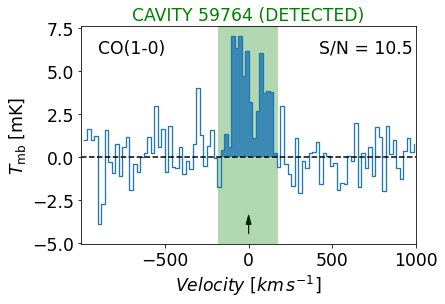}
\includegraphics[width=4.5cm]{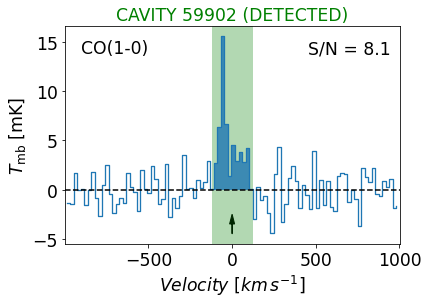}
}
\quad
\centerline{
\includegraphics[width=4.5cm]{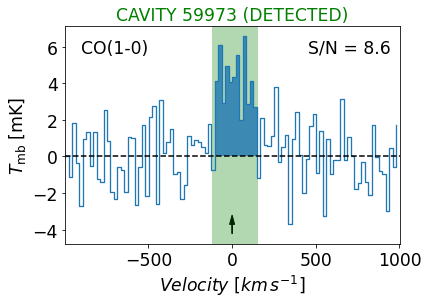}
\includegraphics[width=4.5cm]{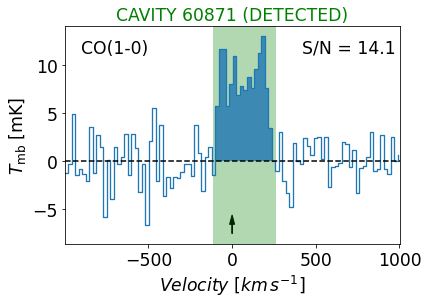}
\includegraphics[width=4.5cm]{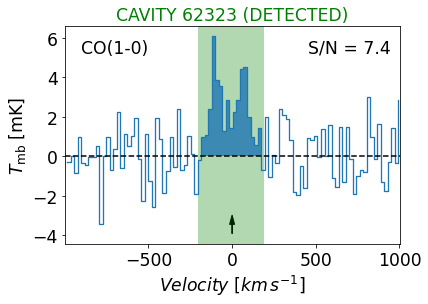}
\includegraphics[width=4.5cm]{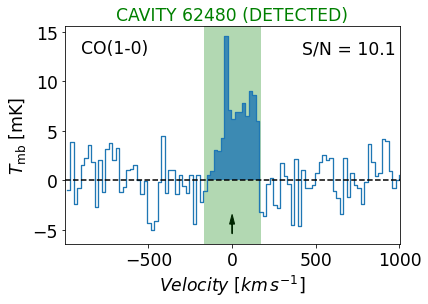}
}
\quad
\centerline{
\includegraphics[width=4.5cm]{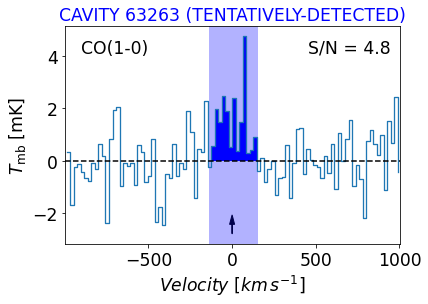}
\includegraphics[width=4.5cm]{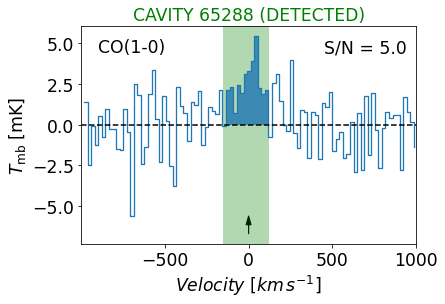}
\includegraphics[width=4.5cm]{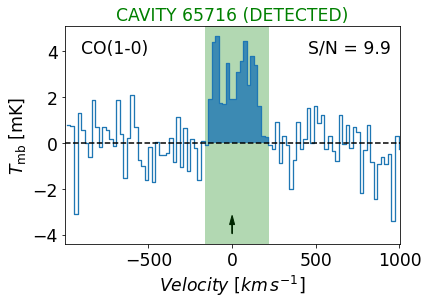}
\includegraphics[width=4.5cm]{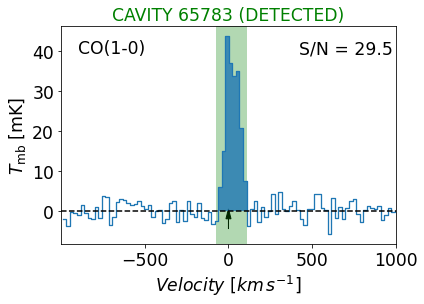}
}
\quad
\centerline{
\includegraphics[width=4.5cm]{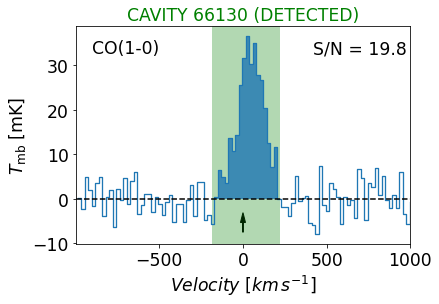}
}
\addtocounter{figure}{-1}
\caption{(continued)}
\end{figure*}


\begin{figure*}
\label{fig:spectra-co21}
\centerline{
\includegraphics[width=4.5cm]{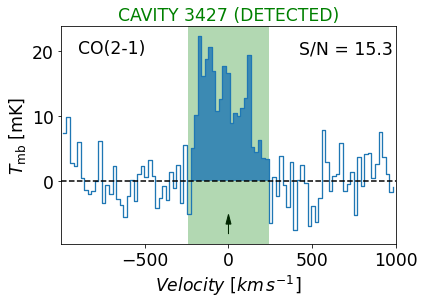}
\includegraphics[width=4.5cm]{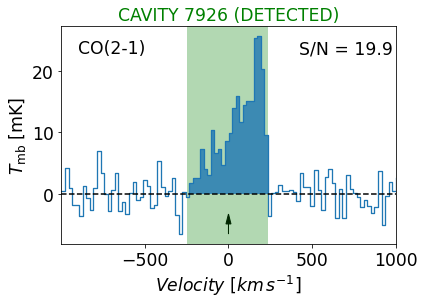}
\includegraphics[width=4.5cm]{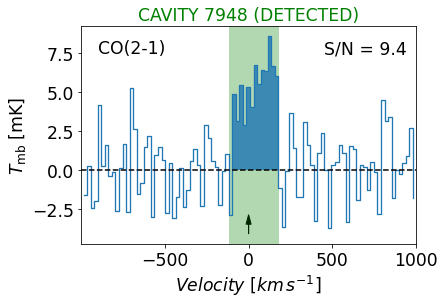}
\includegraphics[width=4.5cm]{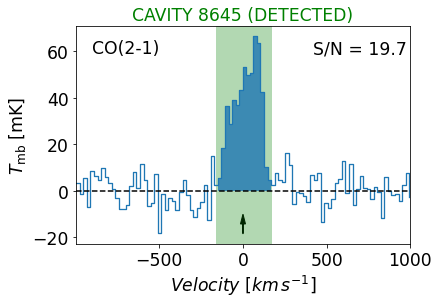}
}
\quad
\centerline{
\includegraphics[width=4.5cm]{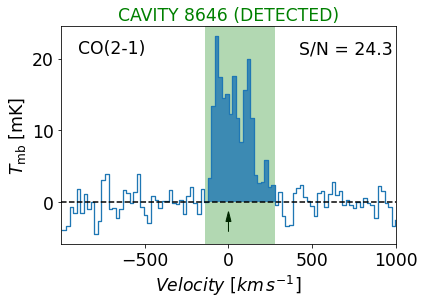}
\includegraphics[width=4.5cm]{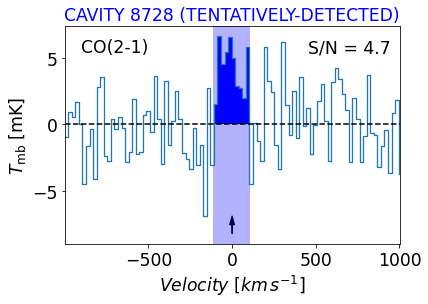}
\includegraphics[width=4.5cm]{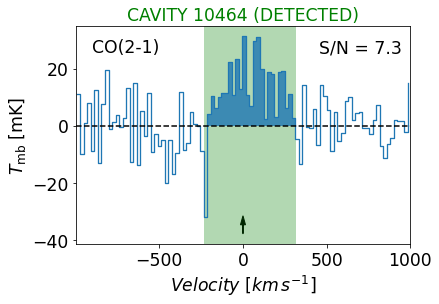}
\includegraphics[width=4.5cm]{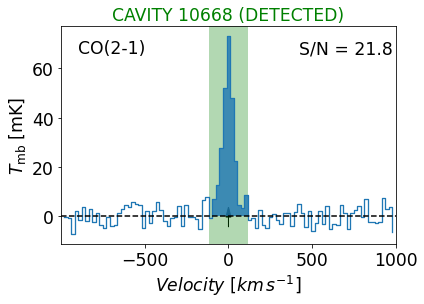}
}
\quad
\centerline{
\includegraphics[width=4.5cm]{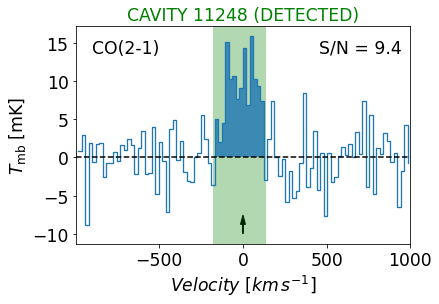}
\includegraphics[width=4.5cm]{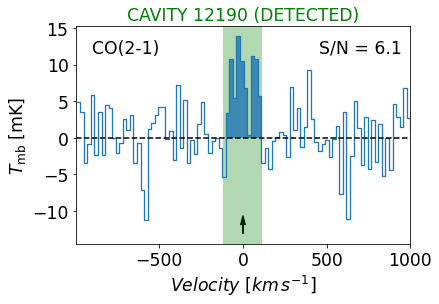}
\includegraphics[width=4.5cm]{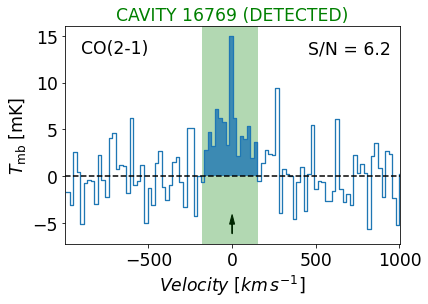}
\includegraphics[width=4.5cm]{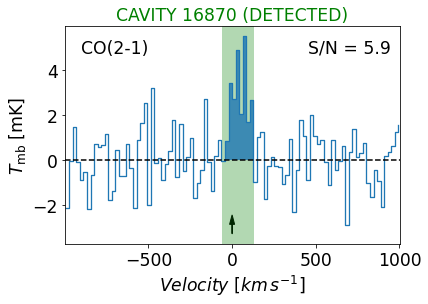}
}
\quad
\centerline{
\includegraphics[width=4.5cm]{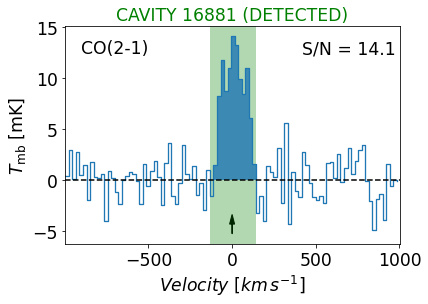}
\includegraphics[width=4.5cm]{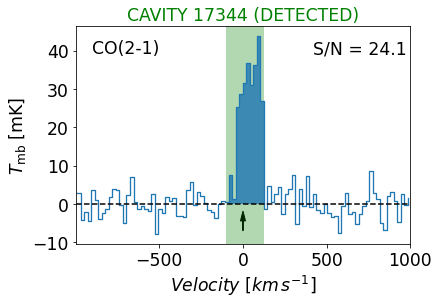}
\includegraphics[width=4.5cm]{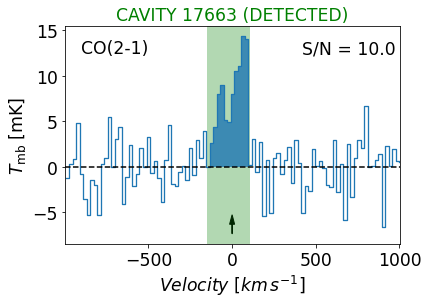}
\includegraphics[width=4.5cm]{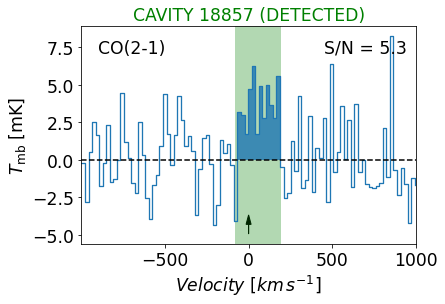}
}
\quad
\centerline{
\includegraphics[width=4.5cm]{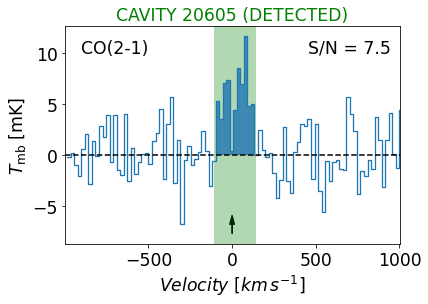}
\includegraphics[width=4.5cm]{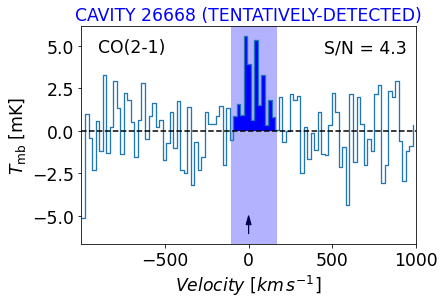}
\includegraphics[width=4.5cm]{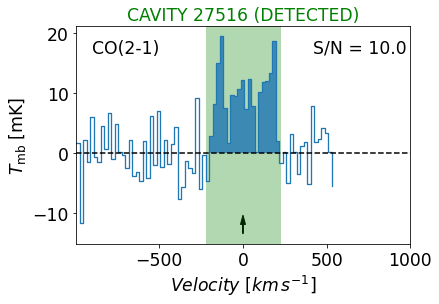}
\includegraphics[width=4.5cm]{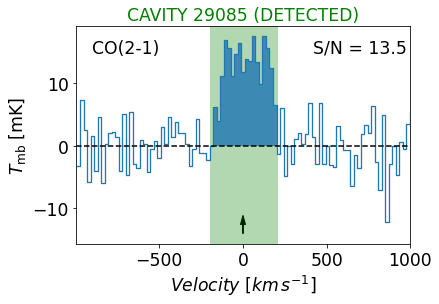}
}
\quad
\centerline{
\includegraphics[width=4.5cm]{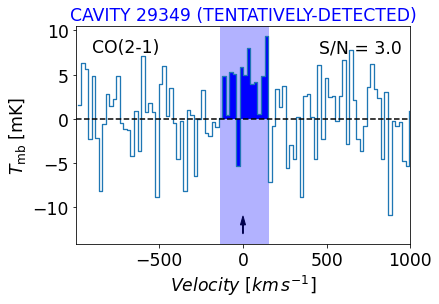}
\includegraphics[width=4.5cm]{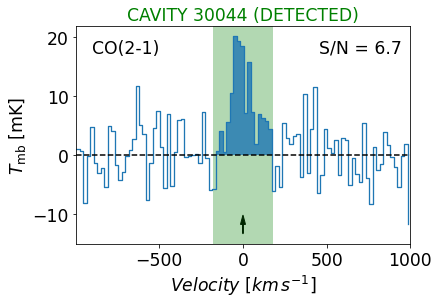}
\includegraphics[width=4.5cm]{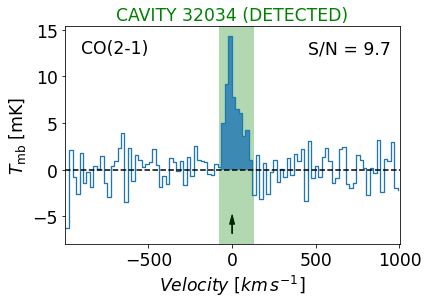}
\includegraphics[width=4.5cm]{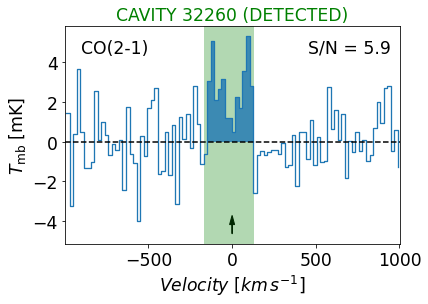}
}
\quad
\centerline{
\includegraphics[width=4.5cm]{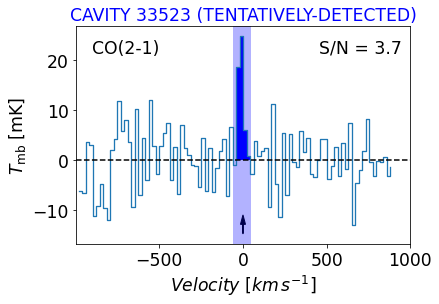}
\includegraphics[width=4.5cm]{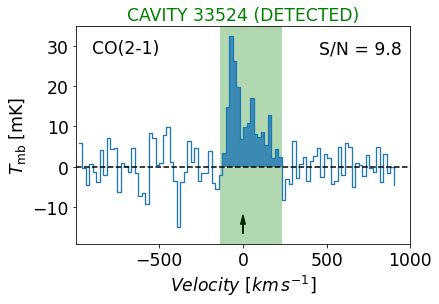}
\includegraphics[width=4.5cm]{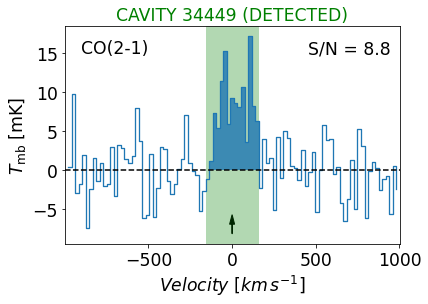}
\includegraphics[width=4.5cm]{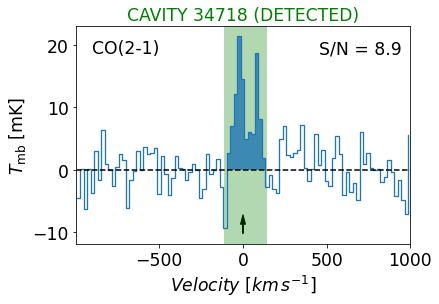}
}
\addtocounter{figure}{0}
\caption{CO(2--1) spectra of the void galaxies. The velocity resolution is 20 km s$^{-1}$. The zero velocity corresponds to the optical redshift $z_{SDSS}$. The coloured shaded area represents the region over which the line is integrated to determine the total flux.}
\label{co21-emission}
\end{figure*}

\begin{figure*}
\centerline{
\includegraphics[width=4.5cm]{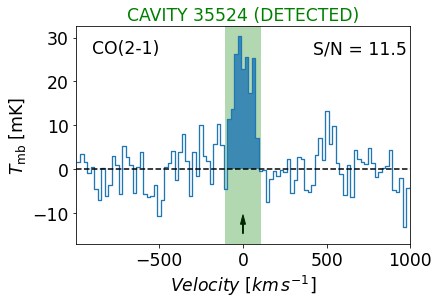}
\includegraphics[width=4.5cm]{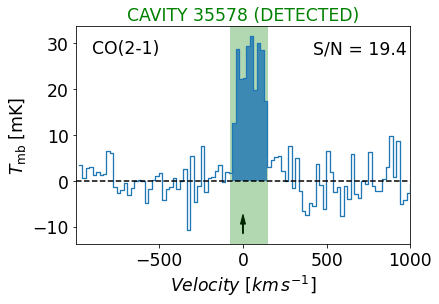}
\includegraphics[width=4.5cm]{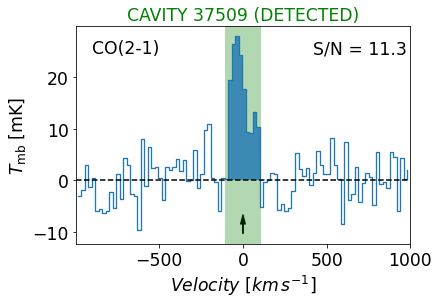}
\includegraphics[width=4.5cm]{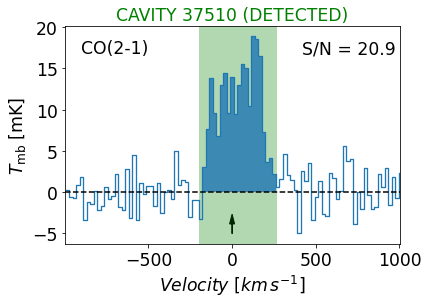}
}
\quad
\centerline{
\includegraphics[width=4.5cm]{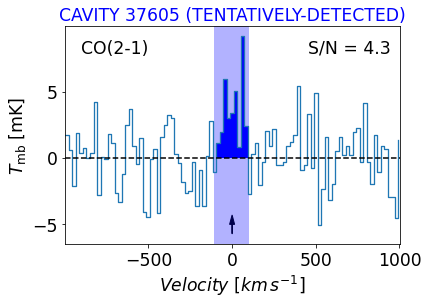}
\includegraphics[width=4.5cm]{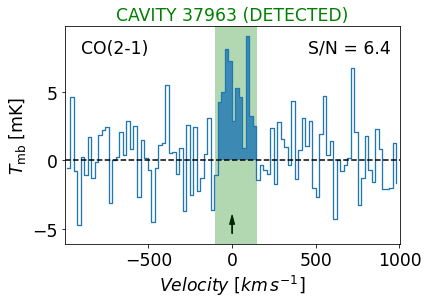}
\includegraphics[width=4.5cm]{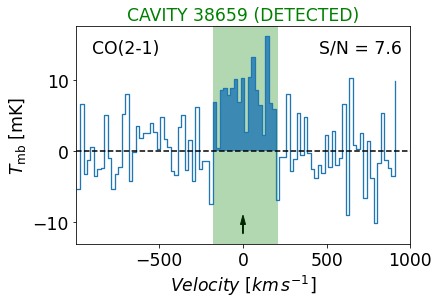}
\includegraphics[width=4.5cm]{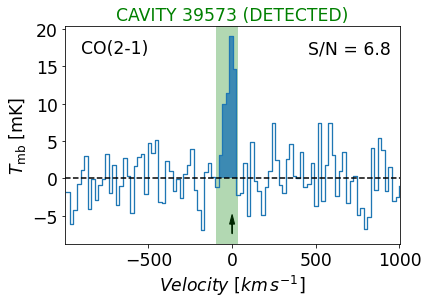}
}
\quad
\centerline{
\includegraphics[width=4.5cm]{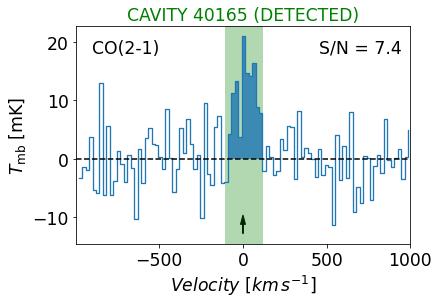}
\includegraphics[width=4.5cm]{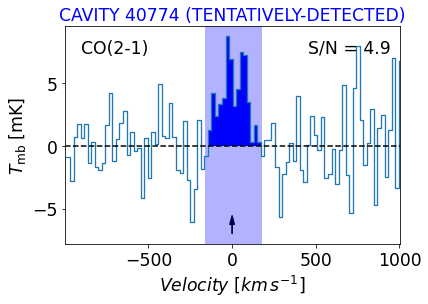}
\includegraphics[width=4.5cm]{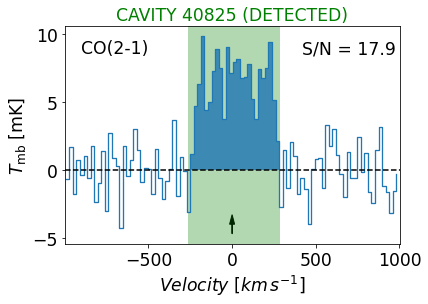}
\includegraphics[width=4.5cm]{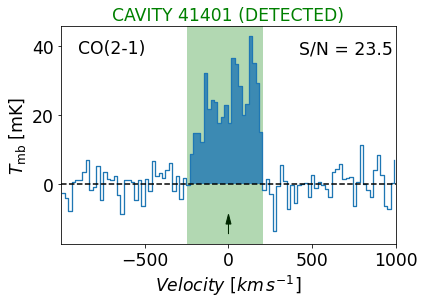}
}
\quad
\centerline{
\includegraphics[width=4.5cm]{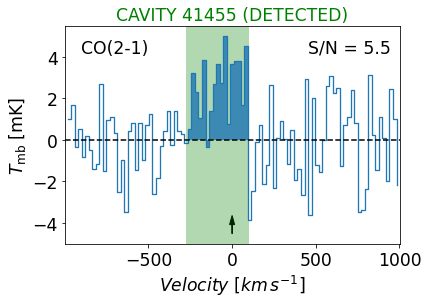}
\includegraphics[width=4.5cm]{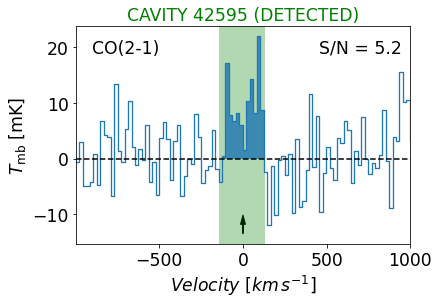}
\includegraphics[width=4.5cm]{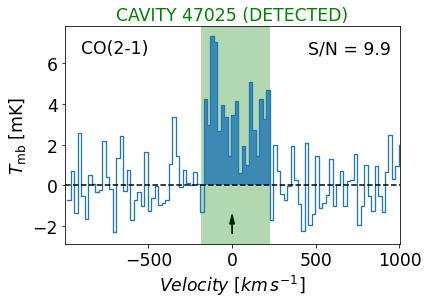}
\includegraphics[width=4.5cm]{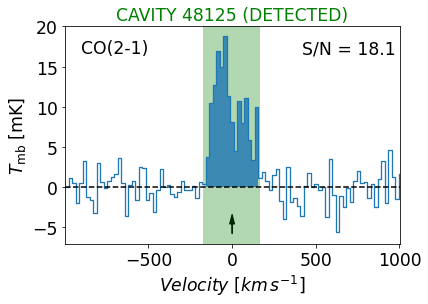}
}
\quad
\centerline{
\includegraphics[width=4.5cm]{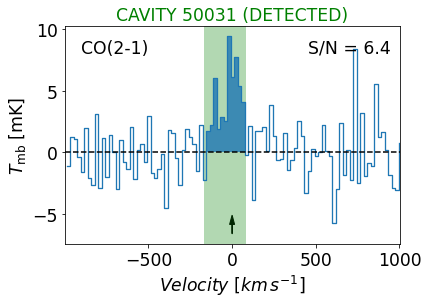}
\includegraphics[width=4.5cm]{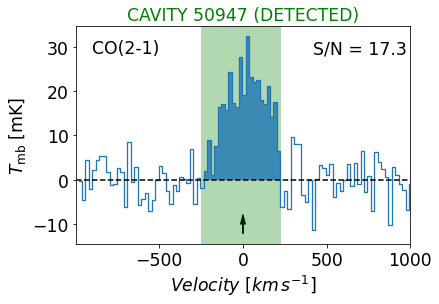}
\includegraphics[width=4.5cm]{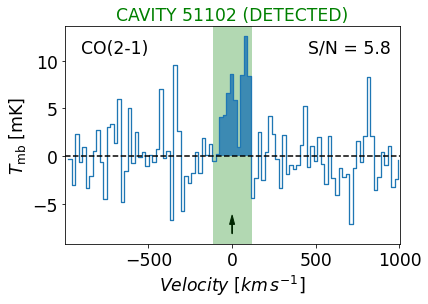}
\includegraphics[width=4.5cm]{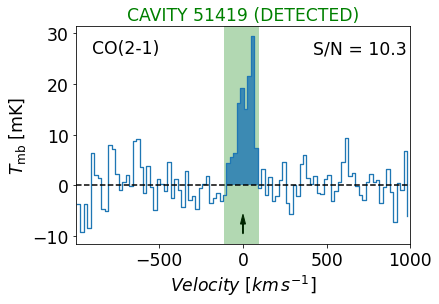}
}
\quad
\centerline{
\includegraphics[width=4.5cm]{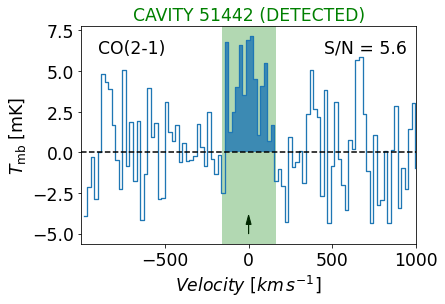}
\includegraphics[width=4.5cm]{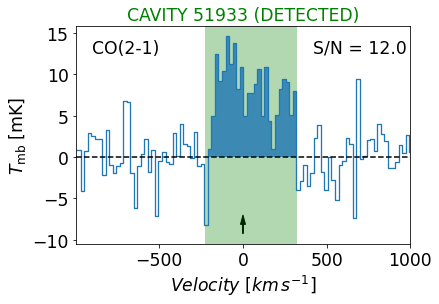}
\includegraphics[width=4.5cm]{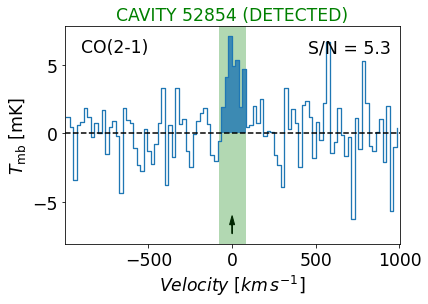}
\includegraphics[width=4.5cm]{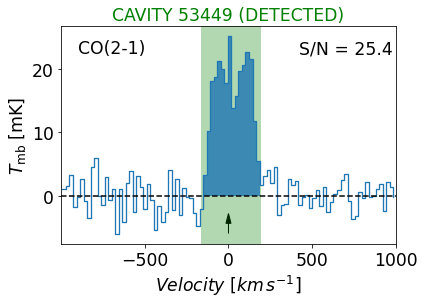}
}
\quad
\centerline{
\includegraphics[width=4.5cm]{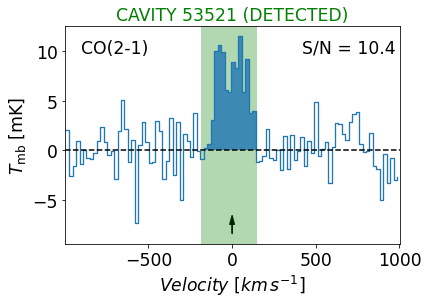}
\includegraphics[width=4.5cm]{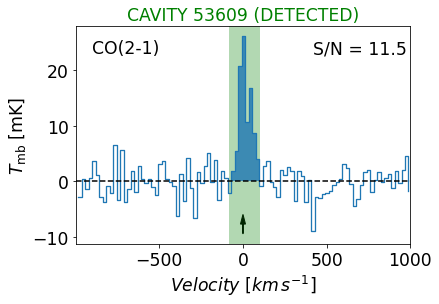}
\includegraphics[width=4.5cm]{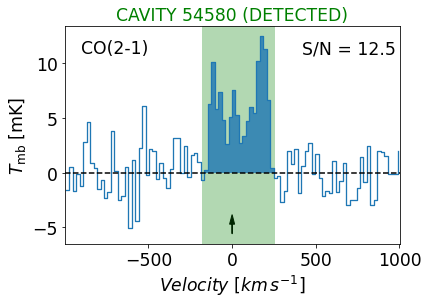}
\includegraphics[width=4.5cm]{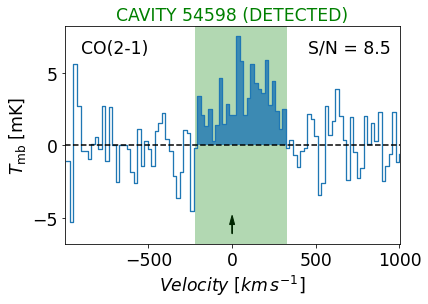}
}
\addtocounter{figure}{-1}
\caption{(continued)}
\end{figure*}

\begin{figure*}
\centerline{
\includegraphics[width=4.5cm]{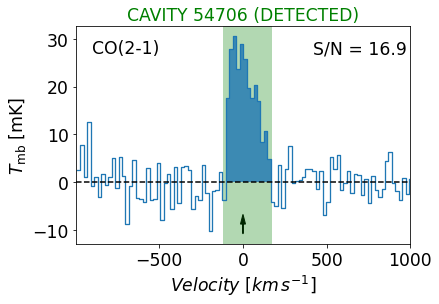}
\includegraphics[width=4.5cm]{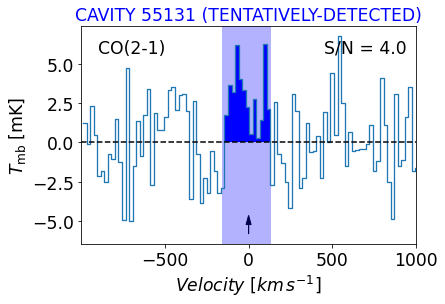}
\includegraphics[width=4.5cm]{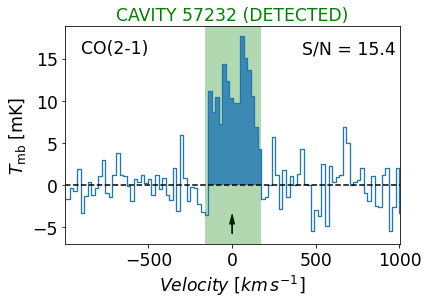}
\includegraphics[width=4.5cm]{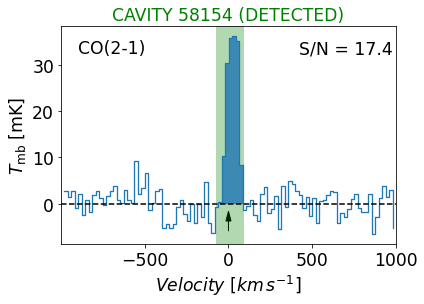}
}
\quad
\centerline{
\includegraphics[width=4.5cm]{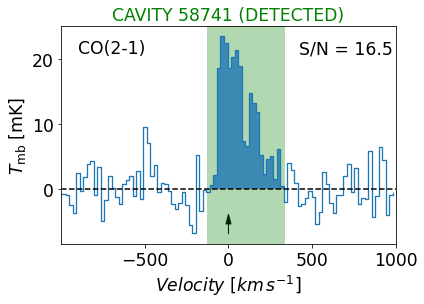}
\includegraphics[width=4.5cm]{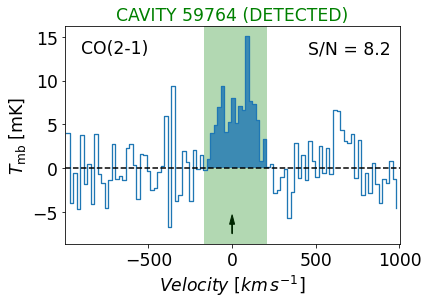}
\includegraphics[width=4.5cm]{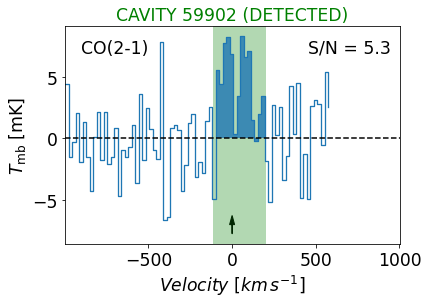}
\includegraphics[width=4.5cm]{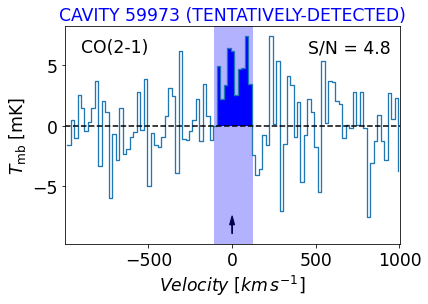}
}
\quad
\centerline{
\includegraphics[width=4.5cm]{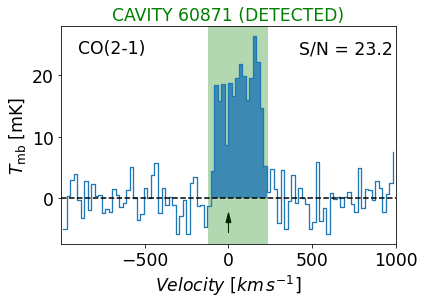}
\includegraphics[width=4.5cm]{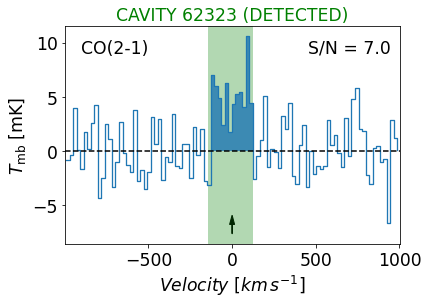}
\includegraphics[width=4.5cm]{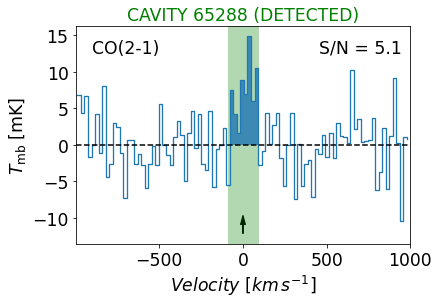}
\includegraphics[width=4.5cm]{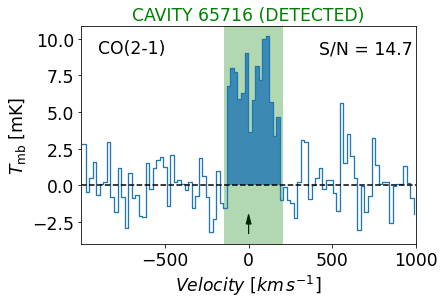}
}
\quad
\centerline{
\includegraphics[width=4.5cm]{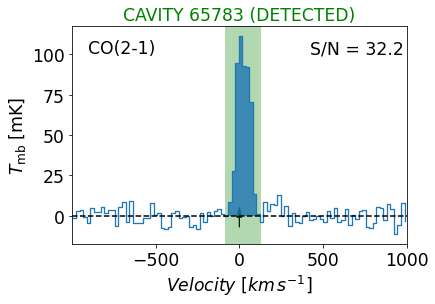}
\includegraphics[width=4.5cm]{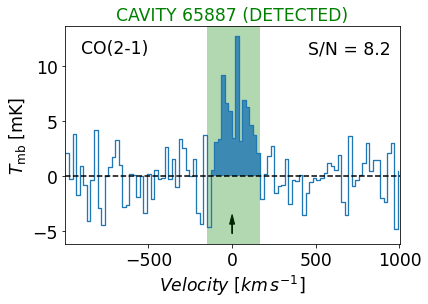}
\includegraphics[width=4.5cm]{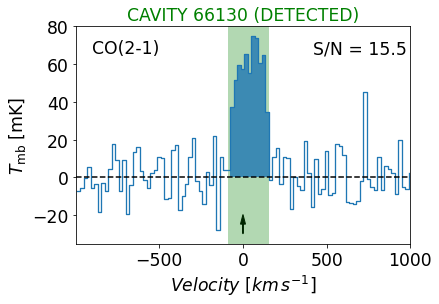}
}
\addtocounter{figure}{-1}
\caption{(continued)}
\end{figure*}


\renewcommand{\thefigure}{B\arabic{figure}}

\section{Comparison of different SFR tracers}
\label{app:SFR}

In order to carry out a reliable comparison between the comparison and void samples, it is crucial to ensure that the parameters used, such as SFR, are consistent despite being calculated using different methods. In the CO-CS sample, SFRs are determined through NUV and MIR emissions from the GALEX and WISE surveys, expressed as SFR$_{\rm NUV+MIR}$ = SFR$_{\rm NUV}$ + SFR$_{\rm W4}$. Here, SFR$_{\rm NUV}$ = 10$^{-28.165}$ L$_{\rm NUV}$ [erg~s$^{-1}$Hz$^{-1}$], 
where L$_{\rm NUV}$ is the NUV luminosity, and SFR$_{\rm w4}$ = 7.50 $\times$ 10$^{-10}$ $\times$ (L$_{\rm w4}$ - 0.044 L$_{\rm w1}$) [L$_{\odot}$], where L$_{\rm w4}$ and L$_{\rm w1}$ are the MIR luminosities in WISE band 4 and WISE band 1, respectively, following the methodology given in \cite{janowiecki17}. On the other hand, for the CO-VG sample, a substantial portion of the sample lacks NUV and/or W4 data, being available for only 151 galaxies. 
Therefore, the SFR  for the CO-VG sample is determined based on H$\alpha$ and H$\beta$ fluxes from the SDSS data, denoted as SFR$_{\rm H\alpha}$, following the procedure detailed in \cite{dp17}. There are a few cases (8 galaxies from the IRAM-30\,m CO-CAVITY survey and 19 from the CO-CS void galaxies subset) for which good-quality H$\alpha$/H$\beta$ lines were not available for estimating the SFR. In these cases we used GALEX and WISE data to calculate SFR$_{\rm NUV+MIR}$  \citep[adhering to the prescription of][]{janowiecki17} or, alternatively, for those cases without NUV and/or W4 data, as SFR$_{\rm MIR}$ = 10$^{-42.70}$ L$_{\rm W3}$ [erg~s$^{-1}$] \citep{leroy19}, where L$_{\rm W3}$ is the MIR luminosity for WISE band 3.

In Figs. \ref{sfr_comparison} and \ref{sfr_comparison_xgass}, we present a comparison of the SFR obtained through different methods for the CO-VG galaxies and the comparison sample respectively. Both figures consist of three panels each, displaying the comparison between SFR$_{\rm H\alpha}$ and SFR$_{\rm NUV+MIR}$ (left), SFR$_{\rm MIR}$ (center), and SFR$_{\rm H\alpha}$(MPA-JHU) (right), which denotes the values obtained from the MPA-JHU catalogue \citep{brinchmann04}. This approach aims to assess whether consistent results can be achieved using different methods. For the comparison sample, we used the SFR$_{\rm NUV+MIR}$ values provided by the xCOLD GASS database. For the CO-VG sample, we recalculated the SFR$_{\rm NUV+MIR}$ following the methodology of \cite{janowiecki17}. For both samples, CO-VG, and the comparison sample, we calculated the SFR$_{\rm H\alpha}$ following the procedure in \cite{dp17}}. Additionally, we calculated SFR$_{\rm MIR}$ using only MIR emission (WISE band 3), as prescribed by \cite{leroy19}. 

We found a strong correlation between SFR$_{\rm NUV+MIR}$ and SFR$_{\rm H\alpha}$. The correlation between SFR$_{\rm H\alpha}$ and  SFR$_{\rm MIR}$ is good, but weaker than the previous one, particularly at low SFR values for the CO-VG sample and at high SFR values for the CO-CS sample. 
Finally, the correlation between  SFR$_{\rm H\alpha}$  and SFR$_{\rm H\alpha-MPA}$ is very good for CO-VG. However, SFR$_{\rm H\alpha}$ is systematically higher than SFR$_{\rm H\alpha-MPA}$ for CO-CS galaxies, especially at lower SFR values. This is in agreement with the previous results of \citet{dp17} and \citet{dominguez22}. In conclusion, the SFR$_{\rm H\alpha}$ tracer appears to yield very similar results to the SFR$_{\rm NUV+MIR}$ tracer (the mean difference between the values of SFR$_{\rm NUV+MIR}$ and SFR$_{\rm H\alpha}$ is 0.01 dex for the CO-VG sample and 0.02 for CO-CS) over almost 3 orders of magnitudes. We therefore use SFR$_{\rm H\alpha}$ for the CO-VG sample and  SFR$_{\rm NUV+MIR}$ for CO-CS \citep[defined as SFR$_{\rm best}$ in][]{janowiecki17}.

\begin{figure*}
\includegraphics[width=\textwidth]{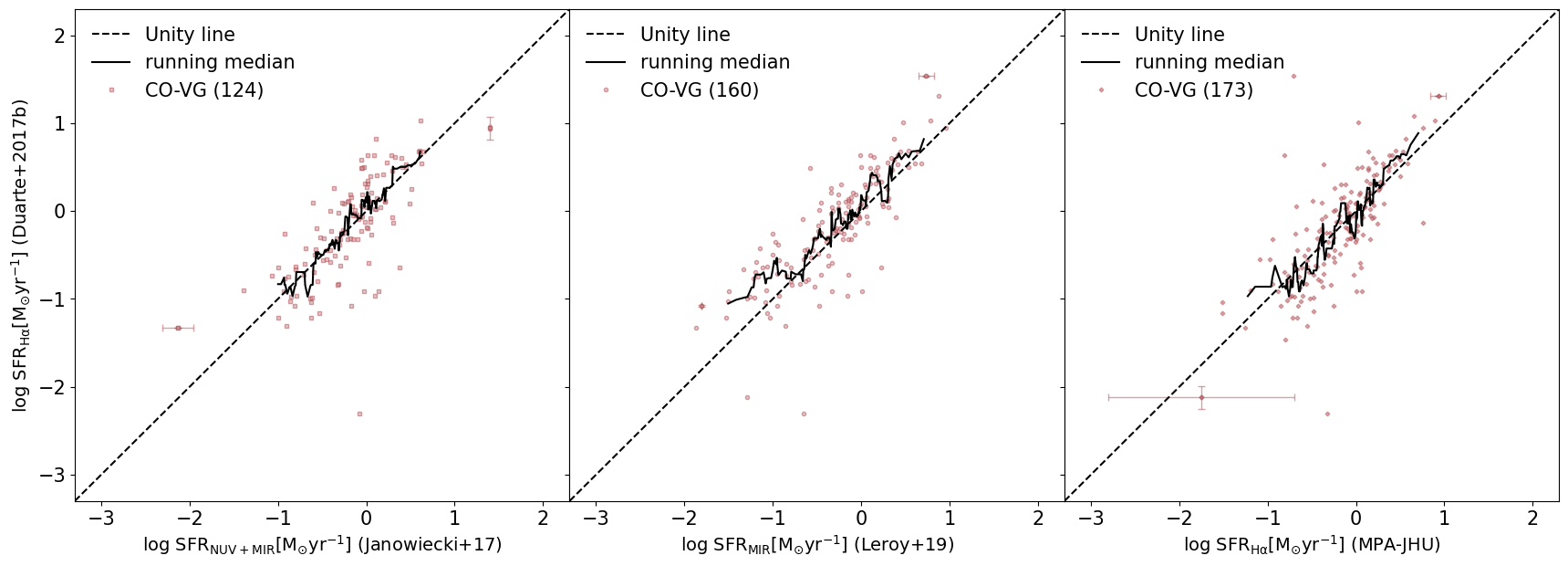}
\caption{Star formation rate tracer (SFR$_{\rm H\alpha}$) derived from H$\alpha$ emission following the method by \cite{dp17} and compared with different SFR tracers for the CO-VG sample. Left: SFR$_{\rm NUV+MIR}$ following the prescription in \cite{janowiecki17}. Centre: SFR$_{\rm MIR}$ following the prescription in \citet{leroy19}. Right: SFR$_{\rm H\alpha-MPA}$ from the MPA-JHU catalogue.}
\label{sfr_comparison}
\addtocounter{figure}{0}
\end{figure*}

\begin{figure*}
\includegraphics[width=\textwidth]{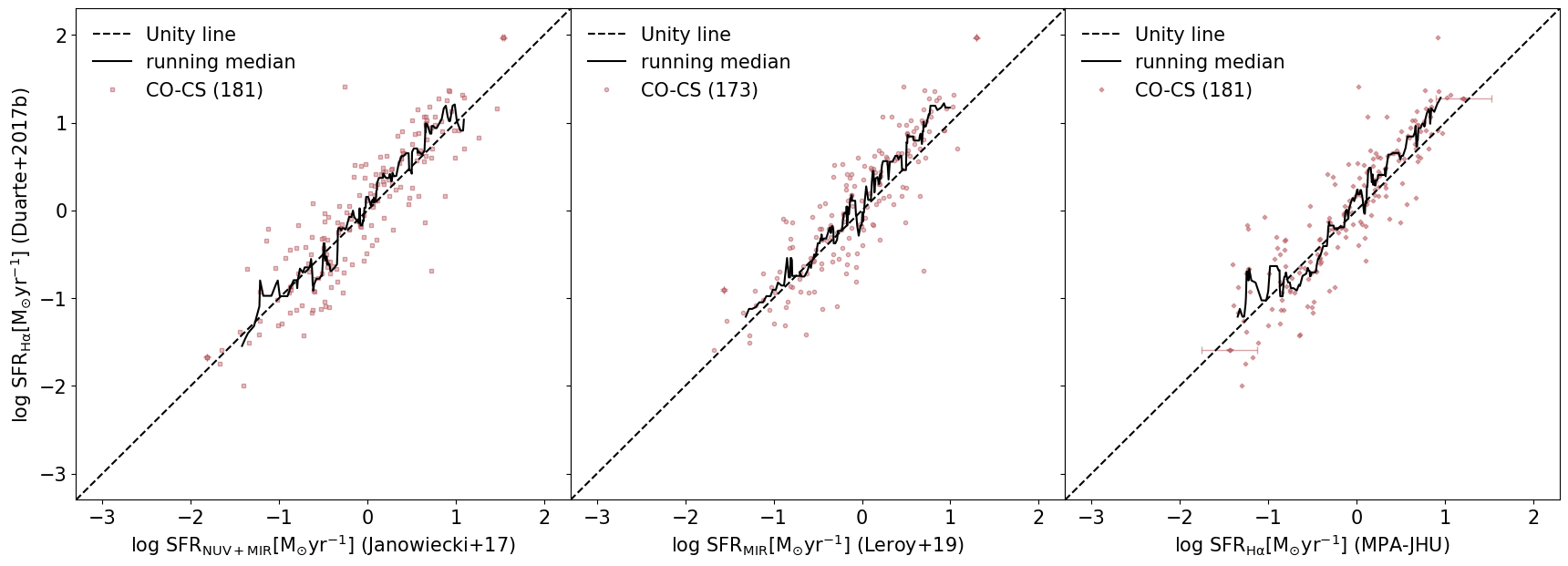}
\caption{Star formation rate tracer (SFR$_{\rm H\alpha}$) derived from H$\alpha$ emission following the method by \cite{dp17} and compared with different SFR tracers for the comparison sample. Left: SFR$_{\rm NUV+MIR}$ following the prescription in \cite{janowiecki20}. Centre: SFR$_{\rm MIR}$ following the prescription in \citet{leroy19}. Right: SFR$_{\rm H\alpha-MPA}$ from the MPA-JHU catalogue.}
\label{sfr_comparison_xgass}
\addtocounter{figure}{0}
\end{figure*}

\end{document}